\documentclass{aa}

\usepackage{graphicx}
\usepackage{txfonts}
\begin{document}

   \title{Cosmic-ray-induced dissociation and reactions in warm interstellar ices}

   \author{J. Kalv\=ans
          }

   \institute{Engineering Research Institute "Ventspils International Radio Astronomy Center" of Ventspils University College, Inzenieru 101, Ventspils, Latvia, LV-3601\\
              \email{juris.kalvans@venta.lv}
             }

   \date{Received September 15, 1996; accepted March 16, 1997}

 
  \abstract
   {Cosmic ray particles that hit interstellar grains in dark molecular cores may induce whole-grain heating. The high temperature of a CR-heated grain allows energy barriers for bulk diffusion and reactions to be overcome. Additionally, ice molecules are destroyed by direct cosmic-ray induced dissociation.}
   {We provide a justified estimate of the significance of cosmic-ray induced surface-mantle diffusion, chemical reactions in ice, and dissociation of ice species in a star-forming interstellar cloud core.}
   {We considered a gas clump in a collapsing low-mass prestellar core and during the initial stages of protostellar envelope heating with a three-phase chemical kinetics model. The model includes a proper treatment of the stochastic aspect of whole-grain heating and new experimental data for dissociation.}
   {We found that the cosmic-ray-induced effects are mostly limited to an increase in abundance for carbon-chain species. The effect on major species' abundances is a few percentage points at most. The HNC:HCN ice abundance ratio in ice is increased.}
   {Among the processes considered in the model, dissociation is probably the most significant, while diffusion and reactions on warm grains are less important. All three processes facilitate the synthesis of complex molecules, including organic species.}


   \maketitle
%

\section{Introduction}
\label{intro}

Dark cores in interstellar molecular clouds are the densest and coldest regions of the interstellar medium. Many such cores are the birth sites of stars. In the dense conditions, most of the chemical elements, heavier than helium, are in the form of dust grains or "dirty ices", which are molecular species frozen onto the grains. Such ices are likely to be chemically processed by high-energy photons, binary reactions, and cosmic rays (CR).

The consequence of a cosmic-ray particle that hits an ice-covered interstellar grain is twofold: (1) a destruction of molecules along the ion's path in the grain and (2) heating of the grain. For the former, whole-grain heating can be considered to be much more important for 0.1$\mu$m grains than local heating \citep{Leger85}. Dissociation may serve as an additional source of chemical radicals in ice, while heating may allow all kinds of energy barriers at molecular level to be more easily overcome. The present study aims to provide a justified estimate about how and to what extent the two processes affect the chemical composition of ices in star-forming regions.

Considerable experimental evidence is available for process (1) -- CR-induced dissociation. Experiments with fast ion irradiation of interstellar ice analogs date back to \citet{Pirronello82} and \citet{Brown82}. Recent estimates indicate that the halftime for icy molecule destruction by cosmic rays in interstellar conditions is several Myr \citep{deBarros11,Pilling10a}, which is probably a few times longer than the lifetime of interstellar ices. A full physico-chemical model is required to investigate the reactions involving radicals produced by dissociation.

Process (2), an elevated temperature for the dust grain, may arise only when a heavy cosmic-ray nuclei (e.g., iron) hits the grain. Such encounters may deliver enough energy to elevate the grain temperature by several tens of K \citep{Leger85}. Evaporation, induced by such whole-grain heating, has been studied before \citep[e.g.,][]{Hasegawa93a}. The effects of enhanced surface diffusion on globally heated grains have been explored by \citet{Reboussin14}. These authors employed a two-phase gas-grain model. \citet{Kalvans14ba} investigated the diffusion of species between the surface and the mantle through cavities in ice.

In the present study, a three-phase model that includes subsurface ice has been used. This provides additional value to the research because bulk reactions can be considered. In particular, cosmic-ray particles are not absorbed by overlying molecules, unlike UV photons. Thus, in the case of ice on interstellar grains, CRs are able to dissociate molecules at all depths. Three-phase models that explicitly consider photoprocessing of subsurface ice mantles are relatively recent \citep{Kalvans10}. Bulk ice chemistry is essential for models that aim to investigate organic molecules in prestellar or protostellar cores \citep{Garrod13}.

The description of the physical and chemical model for a collapsing prestellar and warming protostellar core has been provided in Sects. \ref{phys} and \ref{chem}. It was largely based on the model of \citet[under review]{Kalvans14}. The approach to modeling whole-grain heating-induced chemical processes in interstellar ices is described in Sects. \ref{proc} and \ref{wmod}. CR-induced dissociation modeling is explained in Sect.~\ref{crdissoc}. The results for each of the three processes under consideration -- (1) whole-grain heating-induced radial bulk diffusion, (2) whole-grain heating-induced reactions in ices, and (3) CR-induced dissociation of ice species -- are reviewed in Sect.~\ref{res}. Section~\ref{concl} summarizes the final conclusions.

\section{Methods}
\label{meth}

The model -- `Alchemic-Venta' -- used for this research is an upgraded `ALCHEMIC' code \citep{Semenov10}. The upgrade was carried out by \citet{Kalvans13b,Kalvans14}. In general, the model simulates gas-grain chemistry in a star-forming interstellar cloud core. Special attention was paid to the approach to modeling the processes induced via whole-grain heating by heavy cosmic ray particles. Ices on such grains have been referred to as the `warm' phase, in contrast of the `cold' phase, which refers to ices at thermal equilibrium temperature. It was assumed that all ice species are continuously exchanged between the warm and cold phases. This treats the warm grains as a separate population and ensures that processes on warm grains do not compete with processes on cold grains. More details on this approach are provided in Sect.~\ref{proc}.

\subsection{The physical model}
\label{phys}

\subsubsection{Cloud conditions}
\begin{figure*}
  \vspace{-2cm}
  \hspace{0.5cm}
  \includegraphics[width=18.0cm]{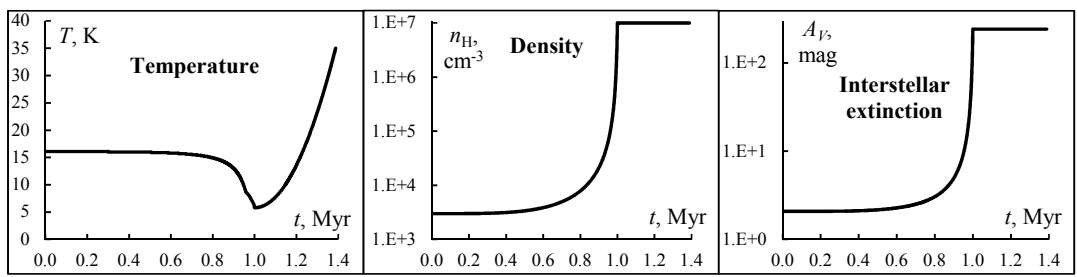}
  \vspace{-20.0cm}
  \caption{Physical conditions in the model.\label{att-phys}}
\end{figure*}
Cosmic rays are a major source of energy in quiescent clouds. This is different for star-forming cores, where gravitational accretion, outflows, fast ions, and X-rays from the protostar affect the nearby nebula.  Isolated low-mass non-starforming cores may be affected by energy input from the interstellar medium. Thus, it is likely that prestellar cores probably are among the objects best suited to investigating CR-induced effects. We considered a contracting prestellar core for the purposes of the present paper. The influence of CRs does not cease with the formation of the protostar. To provide a more detailed insight, a slow warm-up of the circumstellar nebula was modeled for a limited integration time.

The global parameters were chosen with the intent of representing a gas clump in a cloud core that undergoes (1) free-fall gravitational collapse and (2) subsequent heating in the early protostellar envelope. Figure~\ref{att-phys} shows the physical conditions of the core.

For gas density, the approach outlined by \citet{Garrod06} was used. A spherical clump with a mass of 2$M_\odot$ and an initial density $n_{\rm H}=3\times10^3$cm$^{-3}$ collapses isothermally, and the density increases to 10$^7$cm$^{-3}$ over a period of 1Myr. The time-dependent gas density curve was calculated according to \citet{Brown88}. During the second stage, after the supposed formation of the protostar, gas density was retained as constant.

We assumed that a screen of diffuse gas $A_{\rm ext}=1$mag shields the core. The screen was not explicitly considered as part of the collapsing cloud. This means that the extinction in the core is more representative for massive star-formation regions, not isolated globules. The interstellar extinction $A_V$ was calculated self-consistently according to \citet{Savage77}:
   \begin{equation}
   \label{phys1}
A_V = A_\mathrm{ext} + \frac{N_{\rm H + 2H_2}}{1.60 \times 10^{21}}.
   \end{equation}
The total hydrogen column density $N_{\rm H + 2H_2}$ is the number of H atoms (in H{\tiny I} or H$_2$) in a cm$^2$ along the radius of a sphere of uniform density, the prestellar core. The $N_{\rm H + 2H_2}$ was calculated self-consistently for each time integration step. We included the self- and mutual-shielding of H$_2$, CO, and N$_2$ with the use of the tabulated shielding functions \citep{Lee96,Li13}. The required abundances for calculating of column densities of these species were obtained with the chemical model.

We calculated the temperature $T$ of the collapsing core in a time-dependent manner, in line with \citet{Garrod11}. Dust and gas are in thermal equilibrium. In the protostellar stage, an increase in temperature up to 35K was allowed, according to the slow T2 warm-up profile of \citet{Garrod06}. This means that 35K is reached in $3.88\times10^5$ years. This brings the total time of the simulation to 1.388Myr.

The effects of CR-induced warming become increasingly insignificant at higher ice temperatures. For a typical reaction barrier of 1200K, the temperature at which reactions on grains in thermal equilibrium become faster than those on CR-heated grains is around 33K. Additionally, evaporation becomes important and significantly changes ice composition even below this temperature. This further diminishes the CR-induced effects that accumulate over time. To facilitate understanding of the significance of the CR-induced processes, we chose 35K as the final temperature of the simulation run. This is unlike 200K \citep{Garrod06,Garrod08} or 400K \citep{Garrod13}, which was the chosen final temperature for models that study complex organic molecules (COM).

Data from Table 3 of \citet{Garrod06} was used for initial abundances of chemical species in the model. This means that hydrogen is in molecules, while all other elements are in atomic form. The rate of hydrogen ionization by CRs was taken to be $1.3 \times 10^{-17}$s$^{-1}$. The flux for CR-induced photons was taken to be 4875s$^{-1}$cm$^{-2}$ \citep{Roberts07} and that of interstellar photons -- $1.7 \times 10^{8}$s$^{-1}$cm$^{-2}$ \citep{Tielens05}. The last two values were used solely for calculating desorption rates (Sect.~\ref{gasgr}).

\subsubsection{Grain and ice description}
\label{phys-gr}

The model considers grains with a radius of $a=10^{-5}$cm and an additional ice thickness $b$. Here, $b=B\times b_{\rm ML}$, where $B$ is the ice thickness in monolayers, and $b_{\rm ML}=3.7\times10^{-8}$cm is the thickness of a single monolayer (ML). It was assumed that each (rough) grain has $N_s=1.5\times10^6$ adsorption sites on the surface. The number density of the grains is $n_d\approx1.3\times10^{-12}n_{\rm H}$cm$^{-3}$, adopted from the ALCHEMIC model. The maximum ice thickness attained in the model was 173ML at the end of the core contraction epoch. This does not change for the several simulation runs considered in this paper and is very similar to the result obtained by \citet{Kalvans14}.

We employed a three-phase (gas, surface, and ice mantle) chemical model, as proposed by \citet{Kalvans10}, although the ice chemical description is more like that of \citet{Garrod13}. The subsurface ice layer is formed with surface molecules being continuously transferred to the mantle. The rate coefficient for species' transition from the surface to the mantle was assumed to be equal to $t_\mathrm{comp}^{-1}$, where $t_\mathrm{comp}=2 \times 10^{12} \times B_S^2$s. The transfer is halted if surface thickness is lower than 2ML, i.e., if there are fewer than a total of $3 \times 10^6$ species per grain in the surface phase. This approach approximately mimics the total accretion rate of atoms and molecules onto the grain surfaces. It has been derived from studies of compaction of porous ices under the influence of energetic processes \citep{Palumbo06,Kalvans13a}.

The compaction process is intended to describe the ice structure in a time-dependent manner, particularly during the growth of the ice layer on the grains. The slow accretion of free atoms during the early low-density stages of core collapse produces a compact ice with a thin surface layer. \citet{Oba09} demonstrated the formation of a compact structure for ices that form in chemical reactions. When the gas density approaches maximum, the rapid freeze-out of CO and other molecules produces a temporary porous ice, i.e., one with a large surface area. The porosity of ices formed by a fast accretion of inert species has been confirmed by experiments and the rigorous Monte-Carlo modeling method \citep{Chang12}. The experiments \citep{Gerakines96,Palumbo06,Accolla11} have also shown that compaction occurs when ice is exposed to UV photon or ion irradiation, or to exothermic reactions. \citet{Kalvans13a} assume that these processes are the likely cause of the reduction of surface layer thickness after the accretion peak 

In the model, ice compaction is enabled as long as the grains experience net accretion. Molecule exchange between the surface and the mantle layers after this period only occurs via ordinary diffusion. This approach means that species with lower binding energy in the mantle $E_{b,M}$ reach the surface at a faster rate, according to the thermal diffusion-evaporation experiment and model by \citet{Fayolle11} (Sect.~\ref{diff}). Higher order effects, such as the entrapment of volatile species in ice pores and their release during water ice crystallization, were not considered here. This is mostly because a model with a single subsurface layer is poorly suited to detailed modeling of such ice structure-related phenomena.

\subsection{Chemical model}
\label{chem}

The rates of physical transitions and chemical reactions are calculated with the rate equation method. The general first-order rate equation for an unidirectional process $i \longrightarrow j$ is
   \begin{equation}
   \label{chem1}
\frac{\mathrm{d}n_{j,f}}{\mathrm{d}t} = k_1 n_{i,f0},
   \end{equation}
where d$n_{j,f}/$d$t$ is the abundance change (cm$^{-3}$s$^{-1}$) of product species \textit{j} in final phase \textit{f}. $k_1$ is the first-order rate coefficient (s$^{-1}$), and $n_{i,f0}$ is the abundance of reactant \textit{i} in initial phase $f0$ (cm$^{-3}$). The phases \textit{f} and $f0$ are either gas \textit{g}, outer-surface \textit{S}, or mantle \textit{M}. First-order processes include photon- and CR-induced dissociation, and physical phase change processes.

The rate (cm$^{-3}$s$^{-1}$) for a second-order chemical processes (binary reactions) is calculated according to
   \begin{equation}
   \label{chem2}
\frac{\mathrm{d}n_{i,f}}{\mathrm{d}t} = k_{lj,f0}n_{l,f0}n_{j,f0},
   \end{equation}
where \textit{l} and \textit{j} are reactants, \textit{i} is a product molecule, and $k_{lj,f}$ (cm$^{3}$s$^{-1}$) is the second-order rate coefficient.

\subsubsection{Gas and gas-grain interactions}
\label{gasgr}

The chemical reaction network was taken straight from ALCHEMIC as the OSU\_2008\_03 gas-grain database\footnote{Available at \url{http://www.physics.ohio-state.edu/eric/research.html}}. It consists of binary reactions, direct ionization of species by cosmic rays, ionization by cosmic-ray induced photons (grain albedo taken to be 0.5), and ionization by interstellar UV photons. The CR-induced dissociation of ice molecules was included as described in Sect.~\ref{crdissoc}. The rate coefficients of gas-phase reactions ($f0=g$, $f=g$) were calculated exactly as explained in Sect.~2.2. of \citet{Semenov10}. Electron sticking to the neutral grains and the neutralization of atomic ions upon collision with a grain were retained, as described in Sect.~2.3. of \citet{Semenov10}.

It was assumed that reactants that reside in different physical phases (gas, surface, or mantle) never meet. Instead, they may change their phase by several mechanisms. Accretion of gaseous species on grains ($f0=g$, $f=S$) with a sticking coefficient of 0.33 for H and H$_2$, and unity for all other species \citep{Brown90} was considered. The radius of grains $a+b$ was calculated self-consistently in this case, as well as for other processes, i.e., photon-induced desorption and hit rate of CRs. This means that grain size is a time-dependent function with changing parameter $b$ and fixed parameter $a$. Thus, grain growth, arising from the accumulation of ice, was considered in the model, although it is known that it does not significantly influence surface chemistry \citep{Acharyya11,Taquet12}. Evaporation \citep{Hasegawa92} was included for both cold and warm grains at their respective temperature. No accretion or any kind of molecule dissociation was permitted for the warm grains.

Several desorption ($f0=S$, $f=g$) mechanisms were considered for the surface layer. Interstellar and CR-photon-induced desorption was included with yields of 0.003 and 0.002 molecules per photon, respectively. These values have been derived from recent experimental data \citep{Oberg09a,Oberg09b,Bertin13,Fayolle13}. Equal yields for all chemical species were used, which take the codesorption effect into account \citep{Oberg09b,Bertin13}. See \citet{Kalvans14} for a more detailed discussion on photodesorption in the model.

A reaction- and molecule specific reactive desorption was considered, based on the Rice-Ramsperger-Kessel (RRK) theory. This mechanism was studied by \citep{Garrod07} in a general manner, with an uniform efficiency for all species. A similar approach was used in the ALCHEMIC code. In the present study, the heat $E_\mathrm{reac}$ is calculated explicitly for each exothermic reaction with the use of standard enthalpies for products and reactants. The fraction of reactions resulting in product desorption is
   \begin{equation}
   \label{gas1}
f_\mathrm{rd}= \frac{\alpha_\mathrm{RRK} P_\mathrm{RRK}}{1 + \alpha_\mathrm{RRK} P_\mathrm{RRK}},
   \end{equation}
where $\alpha_\mathrm{RRK}$ is a parameter with values 0, 0.01, 0.03, and 0.10 \citep{Garrod07}. The probability $P_\mathrm{RRK}$ for an energy value greater than the desorption energy $E_D$ to be present in the molecule-surface bond is
   \begin{equation}
   \label{gas2}
P_\mathrm{RRK}= \left(1 - \frac{E_D}{E_\mathrm{reac}}\right)^{s-1},
   \end{equation}
where $s=2$ for two-atom species and $s=3N-5$ for other molecules, $N$ the number of atoms in the most complex product molecule, and $E_D$ here is the sum of desorption energies for products of the particular reaction. The assumed value for $\alpha_\mathrm{RRK}$ is 0.03, in line with \citet{Garrod07}.

Finally, indirect reactive desorption was considered. It manifests itself through desorption by the exothermic H+H reactions, which are very common on grains \citep{vandeHulst48,Willacy94}. In a separate study \citep{Kalvans14}, it was found that this mechanism probably fits the requirements for shaping the proportions of major ice components -- water and carbon oxides. A threshold adsorption energy $E_{\rm th}$ \citep{Roberts07} has to be chosen, so that species with higher $E_D$ are not desorbed efficiently, and $E_{D, \mathrm{CO_2}}<E_{\rm th}<E_{D, \mathrm{H_2O}}$. The desorption efficiency $f_{\rm H_2fd}$ for CO and CO$_2$ very likely lies in the range $10^{-6}-10^{-4}$ desorbed molecules per accreted H atom \citep{Kalvans14}. In the present paper, we assumed that $E_{\rm th}=2600$K and $f_{\rm H_2fd}=4\times10^{-6}$ desorbed molecules per accreted H atom. These parameters were chosen to ensure agreement within a factor of two for the abundances of major ice components H$_2$O, CO, and  CO$_2$ observed toward background stars with different interstellar extinction values $A_V$ \citep{Whittet07,Kalvans14}.

\subsubsection{Diffusion processes in ice}
\label{diff}

Diffusion is largely dependent on the binding energies of ice molecules. We assumed that, for the surface layer, molecule binding energy $E_{b,S}=0.5E_D$ \citep{Garrod06}. Similarly, for molecules in the mantle, binding energy was taken to be $E_{b,M}=0.5E_B$, where $E_B=3E_D$ is the assumed absorption energy for species within ice lattice. Absorption energy has been taken significantly higher than $E_D$ with the argument that in ice lattice molecules are surrounded by other species and therefore much more strongly bound to the ice, when compared to molecules on the surface \citep[see also][]{Kalvans14}.

An argument for a high $E_B$ is that $E_{b,M}$ must be higher than $E_D$. This means that evaporation occurs faster than the diffusion of molecules in bulk ice. This is a necessary condition for the ice to be described as a solid. If $E_{b,M}<E_D$, as assumed by \citet{Garrod13}, bulk diffusion (dependent on binding energy in the mantle) is faster than evaporation (dependent on the adsorption energy), which fits the properties of a liquid.

The surface $E_D$ of hydrogen atoms and molecules was modified depending on the surface coverage of H$_2$ molecules ($X_{\rm H_2}$), according to \citet{Garrod11}. Adsorption energy for all other surface species is assumed to be independent of $X_{\rm H_2}$ for the following reasons. The surface thermal-hopping rate ($\approx10^3$s$^{-1}$ for H$_2$ at 10K) is much lower than the rate of vibration for heavier molecules, $\approx10^{12}$s$^{-1}$. (The same also holds true for diffusion via tunneling.) For a surface heavy molecule, which is partially attached to H$_2$, the adjacent hydrogen molecule will very likely soon hop to a different adsorption site. No other species will be able to fill the empty adsorption site, now partially below the heavy molecule, before the latter approaches this site via vibration and binds itself strongly to the ice.

The rate of thermal hopping $R_{\rm hop}$(s$^{-1}$) on the surface was calculated according to \citet{Hasegawa92}, with the use of the respective $E_{b,S}$ for each species. Quantum tunneling for diffusion and reactions were not considered in the model, according to the discussion by \citet{Katz99}. The hopping of molecules in bulk ice was calculated in a similar manner with the use of binding energy $E_{b,M}$. The calculated hopping rates were used for obtaining reaction rate coefficients (Sects. \ref{surfchem} and \ref{manchem}).

The rate of reversible diffusion of molecules between the surface layer and the mantle layer ($f0=S$, $f=M$; and $f0=M$, $f=S$) was calculated as thermal hopping for a molecule in the source layer, until it reaches the target layer:
   \begin{equation}
   \label{dif1}
R_\mathrm{diff}=6R_\mathrm{hop}B_{f0},
   \end{equation}
where $B_{f0}$ is the thickness in MLs of the source layer. The diffusion of a molecule between layers (surface and mantle) involves leaving one layer and making room in another. Considering this, the higher barrier $E_{b,M}$ was used for diffusion in both directions. With the adoption of a cubic geometry for lattice cells, in Eq.~(\ref{dif1}) it was assumed that $6\times B_f0$ hops have to be completed before a molecule leaves its current layer. These diffusion processes were considered, for both, cold grains in the equilibrium temperature, and warm, CR-heated grains. This is not to be confused with diffusion through pores in ice, which was studied by \citet{Kalvans14ba}.

\subsubsection{Surface binary reactions}
\label{surfchem}

Two changes have been made to the OSU\_2008\_03 surface reactions list. The energy barriers of the CO+O$\rightarrow$CO$_2$ and O$_2$+H$\rightarrow$O$_2$H surface reaction were taken to be 290K \citep{Roser01} and 600K \citep{Du12}, respectively.

We calculated the rate coefficient for surface reactions according to \citet{Hasegawa92} and adjusted by reaction-diffusion competition \citep{Garrod11}. Reactions on the discrete grains often occur between species that have been accreted from the gas phase. This may require an approach that takes into account the stochastic aspects of surface chemistry \citep{Caselli98}. The modified-rate equations method was used, as implemented originally in the ALCHEMIC code \citep{Semenov10}. This method has been shown to produce good agreement with the Monte-Carlo random-walk approach \citep{Vasyunin09}. 

The modified rate-equations approach was not applied for surface reactions on warm grains because no accretion was considered for this temporal phase. Additionally, it has been shown that rate equations produce a good fit to the Monte-Carlo results at temperatures of 50K or higher \citep{Du11}.

\subsubsection{Mantle binary reactions}
\label{manchem}

When considering chemical reactions in the mantle, it is important to note that molecules in bulk ice are effectively immobile. The reason for this is that the mantle binding energy is higher than desorption energy (Sect.~\ref{diff}). Molecules that become mobile in the bulk ice tend to diffuse out to the surface and evaporate; i.e., they are not present in the ice for most of the time.

Among major ice constituents, CO is the most volatile molecule with $E_D=1150$K. The temperature range considered in the model spans from 6 to 35K. With $E_{b,M}=1.5E_D$, CO may require more than 80 years to make a single thermal hop to an adjacent absorption site in the mantle at 35K \citep[according to the formalism of][]{Hasegawa92}. This is 17 orders of magnitude more than the time required to overcome the `reactant proximity' barrier of $E_\mathrm{prox}=0.3E_D$ (see below) for a common barrierless reaction with H or other species.

Because the $E_{b,M}$:$E_D$ and $E_\mathrm{prox}$:$E_D$ ratios are the same for all species, this argument works for all possible reactants. Thus, it can be assumed with a degree of confidence that bulk-ice species can be considered frozen and immobile, when calculating rate coefficients of binary reactions.

For reactions in bulk ice, we employed the same reaction set as for surface chemistry. Molecules in ice lattice cells vibrate with their characteristic frequency $\nu_0$, calculated for the absorption energy $E_B$ \citep[see][]{Hasegawa92}. It is assumed that each molecule on average has $N_c=10$ neighbors that can be reached for chemical reactions. This number depends on the (irregular) cell geometry and size. It can be somewhat lower or higher with no significant consequences. To achieve a sufficient proximity for a reaction, a certain energy barrier, $E_{\rm prox}=0.1E_B$ must be overcome. The barrier was assumed to arise from the adsorption force exerted by other neighbors when the molecule leans towards its would-be reaction partner. Barrier height has been derived from the lateral bonding strength for ice species, 0.1$E_D$, as assumed by \citet{Chang12}. The scanning rate (s$^{-1}$) of the cell surface for a molecule is
   \begin{equation}
   \label{manchem1}
R_c = R_\mathrm{hop}/N_c = \nu_0 \mathrm{exp}(-E_\mathrm{prox}/T)/N_c,
   \end{equation}
and the rate coefficient (cm$^{3}$s$^{-1}$) for a binary reaction in ice mantle lattice cell is
   \begin{equation}
   \label{manchem2}
k_{f,ij}= f_\mathrm{act}(ij) (R_{c,i} + R_{c,j}).
   \end{equation}
Here, $f_\mathrm{act}$ is the reaction probability, whose value was calculated according to \citet{Garrod11}, Eq.~(6). The modified rate-equations approach was not applied to mantle reactions. This method has been developed for surface chemistry, where accretion and desorption is possible. In the mantle, new molecules mostly arrive in dissociation events. The rate coefficients for binary reactions in cold and warm ices are calculated in a similar manner, with the only difference being the temperature.

\subsubsection{Photodissociation of ice molecules}
\label{phdissoc}

Dissociation by interstellar and CR-induced photons are first-order processes (Eq.~\ref{chem1}). We assumed the dissociation properties for gas-phase and ice species to be similar \citep{Ruffle01a}. For subsurface mantle species, the incoming photon flux is attenuated by absorption in the above monolayers. Each ML has an absorption probability of $P_\mathrm{abs}=0.007$ \citep{Andersson08}. The attenuation factor was calculated self-consistently according to
   \begin{equation}
   \label{phdissoc1}
A_\mathrm{ice}= (1-P_\mathrm{abs})^{B_S+0.5B_M},
   \end{equation}
where $B_S$ is the current number of monolayers in the surface and $B_M$ the number of MLs in the mantle. This means that, for subsurface species, the dissociation rate for the middle monolayer of the mantle was used.

\subsection{Modeling cosmic-ray induced processes in ice}
\label{proc}

The studies of cosmic-ray effects on interstellar ices have been mostly limited to evaporation from grains heated by cosmic rays \citep{Leger85}. For whole-grain heating process, it was assumed that after a hit by iron CR nuclei (Fe-CR), the grain reaches a maximum temperature $T_{\rm cr}=70$K at which it spends a fraction of time $f(70\rm K)$. Most of the evaporation occurs at $T_{\rm cr}$. This approximation was introduced by \citet{Hasegawa93a} and subsequently used in most papers that consider CR-induced whole-grain heating. These basic physical parameters were also retained for the present study.

An accepted approach is to calculate the rate coefficient for CR-induced desorption with the equation
   \begin{equation}
   \label{proc1}
k_\mathrm{cr.des} = k_\mathrm{evap,70} f(70\rm K),
   \end{equation}
where $k_\mathrm{evap,70}$ is the evaporation rate coefficient at 70K, and
   \begin{equation}
   \label{proc2}
f(70 \mathrm{K}) = \pi (a+b)^2 F_\mathrm{cr} t_\mathrm{cr},
   \end{equation}
where $a=10^{-5}$cm is grain radius, $b$,~cm is the thickness of ice layer on grains (Sect.~\ref{phys-gr}), $t_\mathrm{cr} \approx 10^{-5}$s is the time spent at 70K after each hit, and $F_\mathrm{cr}=2.06 \times 10^{-3}$cm$^{-2}$s$^{-1}$ is the adopted Fe-CR flux \citep{Roberts07}.

The above approach is not well suited to the present study. In theory, reactions in warmed-up ice can be implemented with the simple time-averaged method outlined in Eqs.~(\ref{proc1}) and (\ref{proc2}). However, a chemical reaction is a complex process that involves the diffusion of both partners, and their presence in neighboring adsorption sites on the surface (or absorption sites in bulk ice) for a long enough time to allow overcoming the reaction barrier. It is important to note that processes on cold grains in thermal equilibrium do not occur simultaneously and do not compete with processes on CR-heated grains. This aspect has not been considered in chemical models as yet. The details of the method used here are explained in the next section.

An important parameter is the temperature of the warm grains. The basic approach was employed here and used by \citet{Hasegawa93a} and other authors. This means that $T_{\rm cr}=70$K and $t_{\rm cr}=10^{-5}$s. In reality, the temperature of the population of globally heated grains follows a distribution. This is defined, first, by the energy the dust grain receives from an incoming cosmic-ray particle and, second, by the subsequent cooling of the grain. Cooling is probably dominated by the evaporation of volatile species \citep{Leger85}, which means that the cooling rate depends on ice chemical composition. The resulting actual temperature distribution of grains affected by CRs is time dependent and can be quite complex to calculate.

This calculation was not performed here. Instead, we focus on investigating the effects of CRs on interstellar ice chemistry with the use of known and tested assumptions, i.e., that a 0.1$\mu$m grain reaches 70K for $10^{-5}$s after a Fe-CR hit \citep{Hasegawa93a}. The complex $T$ distribution was recognized in the aims of this paper to provide an estimate of the significance of ice mixing and reactions, induced by whole-grain heating. It was not intended to reproduce observational results with this model.

The direct dissociation of ice species by CRs has been extensively studied in laboratories. Quantitative data for different types of fast ions (CR analogs) have been obtained on the dissociation cross-sections of many species that are common in interstellar ices. A general finding is that, on timescales relevant for interstellar ices, cosmic-ray induced dissociation affects a few per cent of ice molecules at most \citep[e.g.,][]{deBarros11}. The available data allows to implement CR-induced dissociation into a chemical model in a straightforward way (Sect.~\ref{crdissoc}).

We considered CR-induced dissociation, mantle-surface diffusion in warmed-up grains, and overcoming of reaction activation barrier in three short, separate studies in the results section. CR-induced desorption \citep{Leger85} and surface diffusion of ice species \citep{Reboussin14} are included, but their effects are not specifically analyzed.

\subsection{Warm grain chemical modeling approach}
\label{wmod}

To allow ice species to undergo processes on CR-heated grains with uniform temperature of 70K \citep{Hasegawa93a}, according to Eq.~(\ref{proc2}), they are transferred to a separate `warm phase', or species set, that has its own list of solid-phase reactions and phase transitions. This transfer occurs with a rate coefficient ($s^{-1}$):
   \begin{equation}
   \label{wmod1}
k_\mathrm{warm} = \pi (a+b)^2 F_\mathrm{cr}.
   \end{equation}
The transfer back from the warm to the cold phase has a rate coefficient
   \begin{equation}
   \label{wmod2}
k_\mathrm{cold} = \frac{k_\mathrm{warm}}{f(70 \rm K)}.
   \end{equation}
This approach ensures that the molecules on warm grains exist as a population, which is distinct from the majority of cold species. The approximate proportion of warm grains is equal to $f(70\rm K)\approx10^{-16}$. This is an approximate value because $b$ in Eq.~\ref{wmod1} is a time-dependent parameter. It is higher than the value used by \citet{Hasegawa93a} because of a higher assumed Fe-CR flux (Sect.~\ref{proc}).

The rate-equations method is useful only for statistically large and uniform samples. One has to evaluate whether such an approach in the present study complies with these requirements. With the parameters specified in Sect.~\ref{proc}, it can be estimated that in a cloud core with a mass of 0.1$M_\odot$, there are approximately $10^{28}$ warm grains at any given instant in time. This is certainly a sufficiently large number to assume that a continuous population of warm ice molecules exists even in the smallest prestellar cores. However, the population of warm ice molecules is located on multiple islands, the grains. This is a common problem for grain surface chemistry, as discussed in Sect.~\ref{surfchem}.

\subsection{Cosmic-ray induced dissociation}
\label{crdissoc}
%
\begin{table*}
\caption{Selected experimental results of molecule irradiation by heavy ions, related to interstellar ices. Cross sections and other data are given for species that were part of an initial mixture (the interstellar ice analog) for each experimental setup.}
\label{tab-exp}
\centering
\begin{tabular}{crrrcrrcr}
\hline\hline
No. & Species & Initial & Ion & Energy, & T, K & Cross section & Source & Used in  \\
 &  & mixture &  & MeV &  & $\sigma_d$, cm$^2$ &  & model?  \\
\hline
1 & CH$_3$ & CH$_4$ & $_{56}$Fe$^{22+}$ & 267 & 15 & 4.57E-13 & \cite{Mejia13} & yes  \\
2 & CH$_3$ & CH$_4$ & $_{70}$Zn$^{26+}$ & 606 & 15 & 1.99E-13 & \cite{Mejia13} &    \\
3 & CH$_4$ & CH$_3$OH & $_{70}$Zn$^{26+}$ & 606 & 15 & 9.80E-14 & \cite{deBarros11} &    \\
4 & CH$_4$ & CH$_4$ & $_{56}$Fe$^{22+}$ & 267 & 15 & 6.80E-14 & \cite{Mejia13} & yes  \\
5 & CH$_4$ & CH$_4$ & $_{70}$Zn$^{26+}$ & 606 & 15 & 7.20E-14 & \cite{Mejia13} &    \\
6 & H$_2$O & HCOOH & $_{56}$Fe$^{22+}$ & 267 & 15 & 2.20E-13 & \cite{Andrade13} &    \\
7 & H$_2$O & H$_2$O & $_{58}$Ni$^{13+}$ & 52 & 13 & 1.10E-13 & \cite{Pilling10b} &    \\
8 & H$_2$O & H$_2$O:CO$_2$ 10:1 & $_{58}$Ni$^{13+}$ & 52 & 13 & 1.00E-13 & \cite{Pilling10b} &    \\
9 & H$_2$O & H$_2$O:CO$_2$ 1:1 & $_{58}$Ni$^{13+}$ & 52 & 13 & 1.00E-13 & \cite{Pilling10b} &    \\
10 & H$_2$O & H$_2$O:NH$_3$ 2:1 & $_{58}$Ni$^{13+}$ & 46 & 13 & 2.00E-13 & \cite{Pilling10a} &    \\
11 & H$_2$O & H$_2$O:NH$_3$:CO 5:3:2 & $_{58}$Ni$^{13+}$ & 46 & 13 & 2.00E-13 & \cite{Pilling10a} & yes  \\
12 & NH$_3$ & H$_2$O:NH$_3$ 2:1 & $_{58}$Ni$^{13+}$ & 46 & 13 & 1.30E-13 & \cite{Pilling10a} &    \\
13 & NH$_3$ & H$_2$O:NH$_3$:CO 5:3:2 & $_{58}$Ni$^{13+}$ & 46 & 13 & 1.40E-13 & \cite{Pilling10a} & yes  \\
14 & C$_2$H$_2$ & CH$_4$ & $_{56}$Fe$^{22+}$ & 267 & 15 & 3.60E-14 & \cite{Mejia13} & yes  \\
15 & C$_2$H$_2$ & CH$_4$ & $_{70}$Zn$^{26+}$ & 606 & 15 & 1.70E-14 & \cite{Mejia13} &    \\
16 & C$_2$H$_4$ & CH$_4$ & $_{56}$Fe$^{22+}$ & 267 & 15 & 1.27E-13 & \cite{Mejia13} & yes  \\
17 & C$_2$H$_4$ & CH$_4$ & $_{70}$Zn$^{26+}$ & 606 & 15 & 1.40E-13 & \cite{Mejia13} &    \\
18 & C$_2$H$_6$ & CH$_4$ & $_{56}$Fe$^{22+}$ & 267 & 15 & 1.77E-13 & \cite{Mejia13} & yes  \\
19 & C$_2$H$_6$ & CH$_4$ & $_{70}$Zn$^{26+}$ & 606 & 15 & 1.56E-13 & \cite{Mejia13} &    \\
20 & CH$_2$OH & CH$_3$OH & $_{70}$Zn$^{26+}$ & 606 & 15 & 9.21E-14 & \cite{deBarros11} & yes  \\
21 & CH$_3$OH & CH$_3$OH & $_{70}$Zn$^{26+}$ & 606 & 15 & 1.38E-13 & \cite{deBarros11} & yes  \\
22 & CO & HCOOH & $_{56}$Fe$^{22+}$ & 267 & 15 & 2.10E-13 & \cite{Andrade13} & yes  \\
23 & CO & CH$_3$OH & $_{70}$Zn$^{26+}$ & 606 & 15 & 4.40E-14 & \cite{deBarros11} &    \\
24 & CO & CO$_2$ & $_{58}$Ni$^{13+}$ & 52 & 13 & 1.10E-12 & \cite{Pilling10b} &    \\
25 & CO & CO$_2$:H$_2$O 1:10 & $_{58}$Ni$^{13+}$ & 52 & 13 & 7.30E-13 & \cite{Pilling10b} &    \\
26 & CO & CO$_2$:H$_2$O 1:1 & $_{58}$Ni$^{13+}$ & 52 & 13 & 4.40E-13 & \cite{Pilling10b} & yes  \\
27 & CO & H$_2$O:NH$_3$:CO  5:3:2 & $_{58}$Ni$^{13+}$ & 46 & 13 & 1.90E-13 & \cite{Pilling10a} &    \\
28 & CO & CO & $_{58}$Ni$^{13+}$ & 50 & 13 & 1.00E-13 & \cite{Duarte10a} &    \\
29 & CO & CO & $_{64}$Ni$^{24+}$ & 537 & 13 & 3.50E-14 & \cite{Duarte10a} &    \\
30 & H$_2$CO & CH$_3$OH & $_{70}$Zn$^{26+}$ & 606 & 15 & 6.45E-13 & \cite{deBarros11} &    \\
31 & H$_2$O$_2$ & H$_2$O & $_{58}$Ni$^{13+}$ & 52 & 13 & 1.00E-12 & \cite{Pilling10b} &    \\
32 & H$_2$O$_2$ & H$_2$O:CO$_2$ 10:1 & $_{58}$Ni$^{13+}$ & 52 & 13 & 9.30E-13 & \cite{Pilling10b} & yes  \\
33 & H$_2$O$_2$ & H$_2$O:CO$_2$ 1:1 & $_{58}$Ni$^{13+}$ & 52 & 13 & 6.90E-13 & \cite{Pilling10b} &    \\
34 & CO$_2$ & HCOOH & $_{56}$Fe$^{22+}$ & 267 & 15 & 2.30E-13 & \cite{Andrade13} & yes  \\
35 & CO$_2$ & HCOOH & $_{56}$Fe$^{22+}$ & 267 & 15 & 2.20E-13 & \cite{Andrade13} &    \\
36 & CO$_2$ & CH$_3$OH & $_{70}$Zn$^{26+}$ & 606 & 15 & 2.40E-13 & \cite{deBarros11} &    \\
37 & CO$_2$ & CO$_2$ & $_{58}$Ni$^{13+}$ & 52 & 13 & 1.80E-13 & \cite{Pilling10b} &    \\
38 & CO$_2$ & H$_2$O:CO$_2$ 10:1 & $_{58}$Ni$^{13+}$ & 52 & 13 & (0.7-1)E-13 & \cite{Pilling10b} &    \\
39 & CO$_2$ & H$_2$O:CO$_2$ 1:1 & $_{58}$Ni$^{13+}$ & 52 & 13 & (1.6-2.1)E-13 & \cite{Pilling10b} &    \\
40 & CO$_2$ & CO$_2$ & $_{58}$Ni$^{11+}$ & 46 & 13 & 1.70E-13 & \cite{Duarte09} &    \\
41 & HCOOH & HCOOH & $_{56}$Fe$^{22+}$ & 267 & 15 & $\approx$ 1E-12 & \cite{Andrade13} & yes  \\
42 & O$_3$ & CO$_2$ & $_{58}$Ni$^{13+}$ & 52 & 13 & 1.60E-12 & \cite{Pilling10b} &    \\
43 & O$_3$ & CO$_2$:H$_2$O 1:1 & $_{58}$Ni$^{13+}$ & 52 & 13 & 5.00E-13 & \cite{Pilling10b} & yes  \\
44 & CH$_3$OCHO & CH$_3$OH & $_{70}$Zn$^{26+}$ & 606 & 15 & 5.88E-13 & \cite{deBarros11} &    \\
45 & CO$_3$ & CO$_2$ & $_{58}$Ni$^{13+}$ & 52 & 13 & 9.70E-12 & \cite{Pilling10b} &    \\
46 & CO$_3$ & CO$_2$:H$_2$O 1:1 & $_{58}$Ni$^{13+}$ & 52 & 13 & 3.10E-12 & \cite{Pilling10b} &    \\
47 & H$_2$CO$_3$ & CO$_2$:H$_2$O 1:1 & $_{58}$Ni$^{13+}$ & 52 & 13 & 9.90E-13 & \cite{Pilling10b} &    \\
\hline
\end{tabular}
\end{table*}
When a cosmic-ray particle passes through the grain, it causes ionization, which probably is the major cause of molecule destruction \citep{Iza06}. A heated cylinder forms, and the heat is then dissipated across the grain \citep{Leger85}. The latter effect produces whole-grain heating. Two components of the cosmic-ray flux were considered in the model. Iron nuclei effectively heat the grains and dissociate ice species. Lighter CR nuclei are able to dissociate ice species at a rate of approximately one third of that for Fe-CR \citep{deBarros11} and are assumed to be unable to significantly heat up the grains.

The first-order (Eq.~\ref{chem1}) rate coefficient for the dissociation of molecule \textit{i} by cosmic-ray particles is
   \begin{equation}
   \label{crdis1}
k_{\mathrm{crd},i} = \sum{\int{\Phi(E)}\sigma_{d,i}(E)\mathrm{d}E},
   \end{equation}
where \textit{E} is the kinetic energy of the projectile; $\Phi(E)=\Phi_{\rm Fe}(E)+\Phi_{\rm li}(E)$ is the flux density for ions between $E$ and $E+\mathrm{d}E$ (MeV$^{-1}$, iron and lighter nuclei, respectively); and $\sigma_{d,i}$ (cm$^{-2}$) is the dissociation cross section for the molecule being considered \citep{deBarros11}.

\subsubsection{Dissociation cross sections}
\label{crossect}

The dissociation cross section is related to the electronic stopping power $S_e$ \citep{deBarros11}:
   \begin{equation}
   \label{crdis2}
\sigma_{d,i} = a S_e^{3/2}.
   \end{equation}
Here, $S_e$ can be calculated with the SRIM code \citep{Ziegler85}, while $\sigma_{d,i}$ is determined experimentally for a number of species, as shown in Table~\ref{tab-exp}. The parameter \textit a is assumed to be independent of \textit E over the whole energy range.

For practical purposes, the cross section for Fe ions has been considered in detail. For species with experimental data available, $\sigma_{d,\rm Fe}(E)$ was obtained as follows. The stopping power for iron ions over the range $10^{-3}$ to $10^3$MeV was calculated with SRIM for pure solid species with an assumed density of 0.8g~cm$^{-3}$. The value of \textit a was obtained by Eq.(\ref{crdis2}) with $\sigma_{d,k}$ specified in Table~\ref{tab-exp}. Then, $\sigma_{d,i}(E)$ was calculated for $10^{-3}$ to $10^3$MeV. When multiple results for experimental cross sections were available, values for ices with atomic composition close to interstellar ices and Fe impactor ions with energies close to 100MeV (where CR flux density is highest) were chosen. In equivalent cases, the higher $\sigma_{d,i}$ value was adopted.

To obtain comparable data for species with no experimentally determined dissociation cross sections, a similar procedure was applied. Fictitious cross sections were adopted for $E=275$ MeV. For species containing up to two atoms of elements from the second row of the periodic table (C, N, O) the fictitious cross section $\sigma_{d,\rm Fe}$(275MeV) was assumed to be $10^{-13}$ cm$^2$. If a compound contains three or more of these atoms, the cross section was assumed to be $10^{-12}$ cm$^2$. In this regard, any third or higher period atoms (Na, Mg, Si, P, S, Cl, Fe) were treated as equal to two second-period atoms. This means that, for example C$_2$H$_5$ and SiH have an assumed $\sigma$ of $10^{-13}$cm$^2$, while $\sigma_d = 10^{-12}$cm$^2$ for OCN and H$_2$CS. These cross-section values were estimated from experimental data with 267MeV Fe$^{22+}$ ions (Table~\ref{tab-exp}). With such approximations, the 8 MeV energy difference (267 and 275 MeV) can be considered unimportant. The parameter \textit a was then obtained by Eq.(\ref{crdis2}). An unity compound correction was applied for stopping power calculations for substances not in the SRIM database.

The real $\sigma_d$ values may vary significantly, especially for cases with no experimental data. However, because the aim was to investigate the importance of CR-induced chemical changes in ice, a conservative and limited, albeit imprecise, approach was considered more adequate than not considering these reactions at all.

\subsubsection{Cosmic-ray flux}
\label{crflux}

The cosmic-ray spectrum (cm$^{-2}$ s$^{-1}$ sr$^{-1}$ (MeV/nucl)$^{-1}$) can be described with
   \begin{equation}
   \label{crdis3}
\frac{\mathrm{d}E}{\mathrm{d}n} = \frac{CE^{0.3}}{(E+E_0)^3} ,
   \end{equation}
where $C = 9.24 \times 10^4$ and $E_0=300$MeV \citep{Webber83,Shen04}. This equation was applied for the energy range of 10$^{-3}$ to 10$^3$MeV. To obtain input data for Eq.~(\ref{crdis1}), the adopted abundance of Fe nuclei in cosmic rays was taken to be $7.13 \times 10^{-4}$ relative to H nuclei \citep{Meyer98}.

\subsubsection{General considerations on CR-induced dissociation}
\label{crgen}

The dissociation of ice molecules by CR particles, other than Fe nuclei, was considered in a much more general manner. It was assumed that these lighter nuclei deliver 25 \% of the CR dissociative power \citep{deBarros11}. Consequently, the adopted molecule dissociation rate coefficient for lighter CR nuclei was
   \begin{equation}
   \label{crdis4}
k_{\rm crd,li} = 0.33 k_{\rm crd,Fe},
   \end{equation}
where $k_{\rm crd,Fe}$(s$^{-1}$) is the rate coefficient for Fe-CR induced dissociation. This was calculated with Eq.~(\ref{crdis1}).

Because cosmic rays consist of different nuclei that can have a range of energies, an extended set of dissociation products for each reaction was deemed desirable. Three hundred nine ice molecule dissociation reactions (i.e., reaction products) were obtained by combining (photo)dissociation reactions from the OSU and UDFA12 \citep{McElroy13} databases. For reactions with more than one dissociation channel, statistical branching ratios were applied.

Light cosmic rays are able to dissociate species in ice but cannot warm up the grain. This means that in a model with processes on warm grains fully considered, three quarters of all CR-induced dissociation products go to the warm phase, while one quarter remains in the thermal equilibrium (cold) phase. The former are supposed to arise from Fe-CR induced dissociation, and the latter from interactions with light cosmic rays. For a model with CR-induced dissociation and without warm-grain chemistry (Sect.~\ref{r-crdis}), it was assumed that all dissociation products remain in the cold phase.

\section{Results}
\label{res}

The general results of the presented model, including major ice components and the chemistry of COMs, are broadly discussed in \citet{Kalvans14}. Several processes have been described in Sect.~\ref{proc} that have not been considered in astrochemical models before. These are (1) CR-induced dissociation; (2) surface-mantle diffusion in ices, warmed-up by a Fe-CR hit; and (3) binary chemical reactions in warmed ices. The results of three simulations that consider these three processes are described in separate sections below. For reference, we used a `base' model, which has whole-grain heating desorption as its only process directly induced by CR particles. The effects of cosmic rays accumulate over time and are most pronounced at the end of the simulation run. However, at the final integration time $t$=1.388Myr, several important species, such as N$_2$, have significantly evaporated and ice composition does not correspond to interstellar conditions. The results are presented in tables for $t$=1.350Myr ($T$=29.6K), showing the calculated abundance relative to hydrogen and its ratio against the reference model.

\subsection{Notes on the reference model results}
\label{r-ref}
\begin{figure}
  \vspace{-5cm}
  \hspace{-1cm}
  \includegraphics[width=11.0cm]{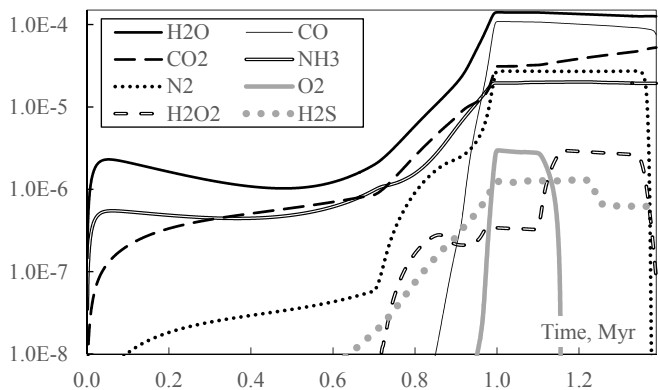}
  \vspace{-5cm}
  \caption{Abundance, relative to hydrogen, of major ice component molecules in the reference model. A very similar picture was obtained for other model runs.\label{att-st}}
\end{figure}
\begin{figure}
  \vspace{-5cm}
  \hspace{-1cm}
  \includegraphics[width=11.0cm]{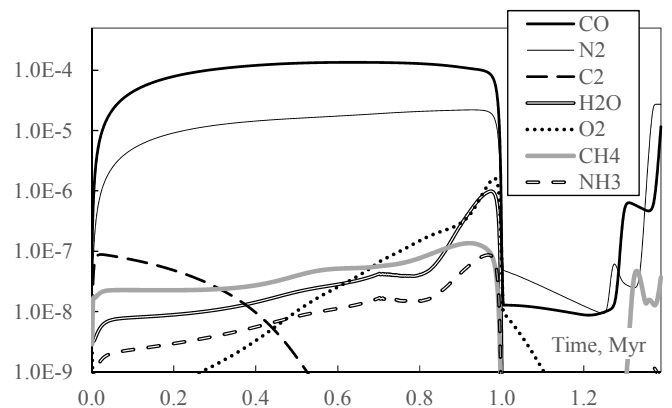}
  \vspace{-5cm}
  \caption{Abundance, relative to hydrogen, of major gas molecules in the reference model. A very similar picture was obtained for other model runs.\label{att-gas}}
\end{figure}
\begin{figure}
  \vspace{-5cm}
  \hspace{-1cm}
  \includegraphics[width=11.0cm]{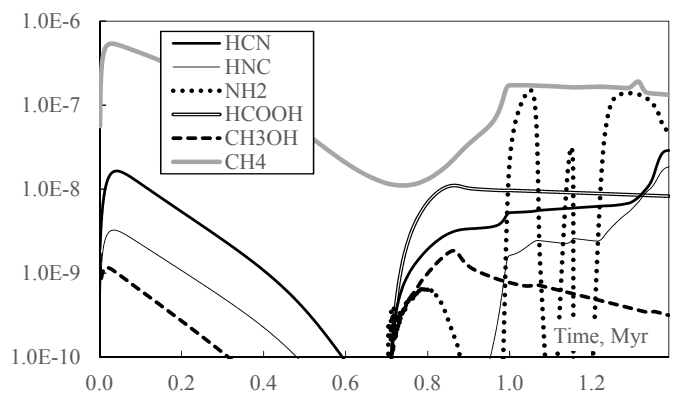}
  \vspace{-5cm}
  \caption{Abundance, relative to hydrogen, of selected ice molecules in the reference model.\label{att-st1}}
\end{figure}

Figures \ref{att-st} and \ref{att-gas} show the relative abundances of major ice and gas-phase molecules. These remain basically unaffected in the various simulations. A discussion is presented here to provide an understanding of mechanisms that govern ice chemistry. I refer the reader to \citet{Kalvans14} for an investigation of the chemistry during the core contraction phase.

The reaction rates in the ice mantle largely depend on the proximity barrier $E_\mathrm{prox}=0.3E_D$, which is specific for each molecule. Species with high adsorption energies react slowly. It turns out that NH$_2$, HS and a few other species are the most important radicals in the subsurface ice, with abundances in excess of $10^{-7}$ relative to hydrogen. NH$_2$ has the highest $E_D$ by far among all major radicals. Among other things, this may result in an efficient synthesis of some nitrogen compounds when the protostar heats up the grains and frozen radicals, such as NH$_2$, become mobile. The reaction network employed here does not include a significant amount of COMs, and the late-stage NH$_2$ daughter species are mostly HCN and HNC, the latter being the major product. The abundance of HNC in ice is doubled during the last 40kyr (1.35-1.39Myr). HNC forms in the barrierless reaction NH$_2$+C. Because of their reactivity and low $E_\mathrm{prox}$, carbon atoms usually have a very low abundance. They are continuously and intensively generated in ice via the photodissociation of CO. Other major photodissociation products do not form stable compounds with NH$_2$.

This result indicates that cyanide and isocyanide are probably rapidly produced in the protostellar warm-up phase. Such a result is not obtained by \citet{Garrod13}, who uses lower $E_{b,M}$ for mantle species and a different approach to mantle reaction rate coefficients. The abundance changes in ice for these, and some other notable species are shown in Fig.~\ref{att-st1}. We refer the reader to \citet{Graninger14} for a recent and detailed investigation of the HCN/HNC chemistry.

Figure~\ref{att-st1} shows that the abundance of NH$_2$ itself is characterized by a series of spikes. Each spike is the result of a separate process. The initial NH$_2$ abundance is low, because the ice layer is thin and mostly confined to the (reactive) surface layer. The first peak around 0.8Myr occurs because NH$_2$ accumulates subsurface in what is mostly a H$_2$O:NH$_3$ ice mixture. NH$_2$ is lost before 0.9Myr, when significant accretion of other species begins (e.g., CO). These species introduce many new radical species via photodissociation that consume NH$_2$ in mutual reactions. At 0.9Myr, $T\approx13$K and $A_V\approx4.7$. These values mean that radicals still are relatively mobile and intensively generated in ice by interstellar photons.

The second peak in NH$_2$ abundance occurs with the massive freeze-out just before 1Myr. The freeze-out peaks at this point because of the rapid increase in density and darkening of the core. Grain temperature falls to 6K and $A_V$ goes into the hundreds. All ice species are practically immobile, and the radicals are only generated by the relatively low flux of CR-induced photons. These are conditions that facilitate a slow accumulation of frozen radicals in ice. The warm-up phase in cloud core evolution begins after 1Myr. When grain temperature nears 8K, atomic radicals become sufficiently mobile, and NH$_2$ is consumed again. When these lighter radicals are depleted, NH$_2$ starts to accumulate again at 1.1Myr. This third peak is made short by the increased mobility of molecular radicals, such as O$_2$H, at 10K.  They create reaction chains that consume NH$_2$.

When these radicals have become depleted, ice abundance of NH$_2$ increases again. It is still generated by the CR-induced photodissociation of ammonia. The final, fourth peak lasts longer and ends, when NH$_2$ and other sticky and heavy radicals themselves become mobile enough to reach their reaction partners in ice lattice cells (speaking in terms of the assumptions of the model).

Other radical species can have one to four peaks, depending on their $E_D$, energy barriers for important sink reactions, and generation from major ice components. Naturally, these curves strongly affect the abundance of minor stable species that are synthesized via these radicals.

Figure~\ref{att-st} shows an example of evaporation (N$_2$) or chemical processing of `stable' molecules (O$_2$, H$_2$O$_2$). The total loss of N$_2$ occurs at a temperature $>$30K, while the minimum evaporation temperature of N$_2$ ($E_D=1000$K) is closer to 20K (evaporation times on the order of 100yr). This is because the loss of a species from bulk ice depends on the ability of these species to diffuse to the surface. This in turn depends on the diffusion energy in the mantle $E_{b,M}$, which is higher than the desorption energy $E_D$.

The transformations of molecular oxygen in the ice mantle can be explained by the gradual increase in temperature. O$_2$ accumulates in large amounts in ice during the freeze-out stage, when $T\approx6$K. In the warm-up phase, when $T\approx10$K, atomic and molecular hydrogen becomes mobile, and O$_2$ is hydrogenated to H$_2$O$_2$. After the temperature exceeds $\approx$30K, further hydrogenation becomes possible with the H$_2$O$_2$+H reaction (barrier 1400K), and hydrogen peroxide is transformed to water.

Aside from atomic species and H$_2$, the main molecules that form in the gaseous phase are CO, N$_2$, and C$_2$. As the cloud contracts, surface chemistry becomes more and more important, and species that form on grain surfaces become abundant. These include mostly H$_2$O, O$_2$, and NH$_3$. In the $\approx$0.4Myr long circumstellar warm-up phase, CO, N$_2$, and CH$_4$, arising from the evaporation of ices, again become highly abundant species. These results are in line with other models of early protostellar cores, e.g. \citet{Garrod06}, \citet{Garrod13}, and \citet{Vasyunin13a}.

\subsection{CR-induced diffusion between the mantle and the surface}
\label{r-wdif}
\begin{table*}
\caption{Calculation results for the model with mantle-surface diffusion in ices, induced by whole-grain heating. Molecular abundances, relative to hydrogen, for selected ice and gas species, and the abundance ratio, relative to the reference model, are shown.}
\label{tab-wdif}
\centering
\begin{tabular}{lcr|lcr|lcr|lcr}
\hline\hline       
\multicolumn{12}{c}{Ice, 1.35Myr}\\
Species & Abund. & Ratio & Species & Abund. & Ratio & Species & Abund. & Ratio & Species & Abund. & Ratio \\
\hline                    
C$_8$H$_2$ & 2.6E-13 & 16.2 & HC$_2$NC & 1.6E-12 & 3.24 & C$_2$H$_4$ & 1.4E-08 & 1.30 & NH$_3$ & 2.0E-05 & 1.03 \\
HC$_7$N & 4.5E-14 & 12.9 & HC$_3$N & 3.8E-10 & 2.87 & CH$_3$NH$_2$ & 4.2E-08 & 1.23 & H$_2$S & 6.3E-07 & 1.01 \\
C$_6$H$_6$ & 3.8E-13 & 12.9 & HNC$_3$ & 1.8E-13 & 2.55 & O$_3$ & 2.4E-12 & 1.22 & H$_2$O & 1.3E-04 & 1.00 \\
CH$_3$C$_4$H & 2.9E-12 & 9.62 & CH$_3$CHO & 2.7E-13 & 2.29 & HNC & 1.4E-08 & 1.20 & CO$_2$ & 4.8E-05 & 0.999 \\
CH$_3$C$_5$N & 1.5E-13 & 7.98 & CH$_3$CN & 3.3E-09 & 1.92 & NH$_2$ & 1.2E-07 & 1.16 & CO & 9.0E-05 & 0.994 \\
C$_6$H$_2$ & 1.3E-11 & 6.80 & H$_2$C$_3$O & 6.3E-12 & 1.83 & OCS & 1.7E-08 & 1.14 & HNO & 7.7E-08 & 0.988 \\
C$_4$H$_4$ & 1.5E-11 & 6.70 & H$_2$CCO & 2.1E-11 & 1.67 & CS & 2.2E-09 & 1.13 & SO & 1.8E-07 & 0.985 \\
HC$_5$N & 2.8E-12 & 5.35 & CH$_3$OCH$_3$ & 3.2E-12 & 1.60 & CN & 2.2E-10 & 1.13 & N$_2$ & 2.6E-05 & 0.985 \\
CH$_3$C$_3$N & 2.9E-11 & 3.92 & C$_2$H$_5$OH & 3.2E-12 & 1.60 & SO$_2$ & 6.1E-10 & 1.13 & NS & 1.8E-08 & 0.984 \\
C$_2$H$_2$ & 9.0E-10 & 3.70 & HNCO & 1.4E-09 & 1.48 & HCN & 1.6E-08 & 1.12 & H$_2$CS & 3.0E-08 & 0.974 \\
C$_4$H$_2$ & 9.8E-10 & 3.55 & C$_2$H$_6$ & 3.9E-08 & 1.40 & O$_2$ & 6.1E-10 & 1.09 & S$_2$ & 1.8E-12 & 0.841 \\
\hline
\multicolumn{12}{c}{Gas, 1.388Myr}\\
H$_2$CCO & 3.5E-12 & 2.67 & C$_3$H$_2$ & 1.8E-12 & 1.95 & CH$_4$ & 6.1E-08 & 1.65 & HNC & 3.2E-10 & 1.28 \\
H$_2$CO & 4.5E-11 & 2.26 & CH$_3$NH$_2$ & 1.2E-12 & 1.95 & C$_2$H$_2$ & 9.6E-11 & 1.54 & HCN & 5.2E-10 & 1.28 \\
H$_2$S & 1.4E-12 & 2.14 & CO$_2$ & 1.7E-08 & 1.87 & C$_2$N & 1.1E-11 & 1.50 & HNO & 3.8E-11 & 1.17 \\
C$_3$ & 1.1E-11 & 2.02 & N$_2$O & 1.6E-12 & 1.86 & C$_2$H$_6$ & 2.1E-12 & 1.48 & CO & 1.2E-05 & 1.04 \\
OCS & 1.1E-12 & 2.01 & H$_2$O & 1.7E-11 & 1.71 & NO & 1.5E-11 & 1.33 & O$_2$ & 5.6E-10 & 0.890 \\
\hline                  
\end{tabular}
\end{table*}
The mantle-surface diffusion rate coefficient depends on the molecular binding energy in ice, which is rather high, $E_{b,M}=1.5E_D$. Even a short temperature increase to 70K can have an observable effect on the hundreds of kyr long timescale, because the mobility of molecules with $E_{b,M}<1300$K is increased significantly. The basic process is the supply of atomic radicals from the mantle to the surface, where they are able to react quickly. Thus, the overall rate of reactions is increased by moving the would-be reactants to a phase (the surface) where they can rapidly diffuse and find a suitable reaction partner.

Table~\ref{tab-wdif} shows that carbon-chain species can have their abundances increased up to a factor of 16, while only very few of the stable species have their abundances decreased. This is because carbon atoms, produced in CO photodissociation, diffuse to the surface where they can link up in larger molecules. Other radicals diffuse to the surface, too; however, their subsequent surface reactions either produce small molecules that are already abundant, or get involved in larger species with a carbon skeleton.

Unlike for ice, the CR-induced effects do not accumulate over time in the gas phase. Abundance changes are relatively small and can usually be observed only for a short period in time -- when an ice molecule, affected by the CR-induced effect, undergoes intensive desorption. Because of this, the calculation results for gas-phase species are poorly suited to being shown graphically. To provide an impression of the significance of the CR-induced processes, the abundance changes for gas-phase species at the end of the simulation run have been shown in results tables. At a temperature of 35K at 1.388Myr, the evaporation of ice species affects the gas phase more than at any time before.

The lower part of Table~\ref{tab-wdif} shows gas-phase abundances ($>10^{-12}$) and ratio against the reference model for the warm diffusion model at the end of the simulation run. Generally, the abundance of species is affected by evaporation of lighter species (e.g., H$_2$CO, CH$_4$) and non-thermal desorption for species with higher adsorption energies (e.g., H$_2$O, CH$_3$NH$_2$). Species produced directly on the surface are most affected. For example, the atoms of molecular oxygen are more effectively converted into CO, CO$_2$, H$_2$O$_2$, H$_2$O, and other species on the surface, which means that the gas-phase abundance of O$_2$ becomes lower, although the abundance of O$_2$ in (subsurface) ice has grown.

\subsection{Reactions in warmed-up ice}
\label{r-warm}
\begin{table*}
\caption{Calculation results for model with whole-grain heating induced chemical reactions in ices. Molecular abundances, relative to hydrogen, for selected ice and gas species, and the abundance ratio, relative to the reference model, are shown.}
\label{tab-warm}
\centering
\begin{tabular}{lcr|lcr|lcr|lcr}
\hline\hline       
\multicolumn{12}{c}{Ice, 1.35Myr}\\
Species & Abund. & Ratio & Species & Abund. & Ratio & Species & Abund. & Ratio & Species & Abund. & Ratio \\
\hline                    
C$_8$H$_2$ & 2.4E-13 & 15.0 & C$_4$H$_2$ & 9.5E-10 & 3.43 & C$_2$H$_6$ & 3.9E-08 & 1.40 & HCN & 1.6E-08 & 1.11 \\
C$_6$H$_6$ & 3.5E-13 & 12.0 & HC$_2$NC & 1.5E-12 & 3.05 & C$_2$H$_4$ & 1.4E-08 & 1.30 & O$_2$ & 6.1E-10 & 1.09 \\
HC$_7$N & 4.0E-14 & 11.7 & HC$_3$N & 3.6E-10 & 2.71 & CH$_3$NH$_2$ & 4.1E-08 & 1.21 & NH$_3$ & 2.0E-05 & 1.03 \\
C$_7$H$_2$ & 1.6E-12 & 9.55 & CH$_3$CHO & 2.6E-13 & 2.25 & O$_3$ & 2.4E-12 & 1.20 & H$_2$S & 6.3E-07 & 1.01 \\
CH$_3$C$_4$H & 2.7E-12 & 9.02 & CH$_3$CN & 3.2E-09 & 1.84 & HNC & 1.3E-08 & 1.18 & H$_2$O & 1.3E-04 & 1.00 \\
CH$_3$C$_5$N & 1.4E-13 & 7.40 & H$_2$C$_3$O & 6.1E-12 & 1.78 & NH$_2$ & 1.2E-07 & 1.15 & CO$_2$ & 4.8E-05 & 0.998 \\
C$_6$H$_2$ & 1.2E-11 & 6.45 & H$_2$CCO & 2.1E-11 & 1.64 & OCS & 1.7E-08 & 1.13 & CO & 9.0E-05 & 0.994 \\
C$_4$H$_4$ & 1.4E-11 & 6.31 & C$_3$H$_4$ & 1.1E-09 & 1.64 & N$_2$H$_2$ & 1.6E-08 & 1.13 & H$_2$O$_2$ & 2.2E-06 & 0.994 \\
HC$_5$N & 2.6E-12 & 4.95 & CH$_3$OCH$_3$ & 3.3E-12 & 1.62 & CS & 2.1E-09 & 1.12 & N$_2$ & 2.6E-05 & 0.985 \\
CH$_3$C$_3$N & 2.7E-11 & 3.71 & C$_2$H$_5$OH & 3.3E-12 & 1.62 & SO$_2$ & 6.1E-10 & 1.12 & S$_2$ & 1.9E-12 & 0.851 \\
C$_2$H$_2$ & 8.7E-10 & 3.58 & HNCO & 1.4E-09 & 1.46 &  &  &  &  &  &  \\
\hline                  
\multicolumn{12}{c}{Gas, 1.388Myr}\\
H$_2$CCO & 3.3E-12 & 2.55 & CH$_3$NH$_2$ & 1.2E-12 & 1.87 & CH4 & 6.0E-08 & 1.61 & HNC & 3.1E-10 & 1.26 \\
H$_2$CO & 4.3E-11 & 2.18 & CO$_2$ & 1.7E-08 & 1.82 & C$_2$H$_2$ & 9.4E-11 & 1.51 & HCN & 5.1E-10 & 1.25 \\
H$_2$S & 1.4E-12 & 2.09 & N$_2$O & 1.5E-12 & 1.80 & C$_2$H$_6$ & 2.1E-12 & 1.48 & HNO & 3.8E-11 & 1.16 \\
OCS & 1.1E-12 & 1.97 & C$_2$ & 5.5E-12 & 1.65 & C$_2$H$_2$N & 1.7E-12 & 1.37 & CO & 1.2E-05 & 1.04 \\
C$_3$H$_2$ & 1.7E-12 & 1.90 & H$_2$O & 1.7E-11 & 1.64 & NO & 1.5E-11 & 1.31 & O$_2$ & 5.7E-10 & 0.898 \\
\hline                  
\end{tabular}
\end{table*}
A higher grain temperature after Fe-CR hits allows overcoming energy barriers more efficiently for species diffusion, reactant proximity in ice, and activation for reactions. The first two effects makes all reactions more efficient, while the third specifically increases the efficiency of reactions with activation barriers.

The addition of reactions in warmed ices into the model gives basic results that are similar to the mantle-surface diffusion model (above section). Both of these mechanisms enhance reaction rates, resulting in more efficient synthesis pathways that mostly involve radicals. Most of the reactions result in the simple parent species -- CO, CO$_2$, H$_2$O, N$_2$, NH$_3$. However, a growing proportion of C atoms accumulate into organic compounds, many of which contain heteroatoms.

Table~\ref{tab-warm} gives insight into the abundance changes of an assortment of ice and gas chemical species. The similarity of the results for the enhanced ice diffusion and reaction rates in warmed ice indicate that the complexity of ice species increases almost regardless of the exact mechanism, which is responsible for promoting reactivity. The chemical effects of whole-grain heating become significant only after $\approx$1Myr, i.e., when the ice layer is thick and extremely cold, which makes other reaction pathways inefficient.

\subsection{The effect of cosmic-ray induced dissociation}
\label{r-crdis}
\begin{table*}
\caption{
Calculation results for model with CR-induced dissociation of ice species. Molecular abundances, relative to hydrogen, for selected ice and gas species, and the abundance ratio, relative to the reference model, are shown.}
\label{tab-crdis}
\centering
\begin{tabular}{lcr|lcr|lcr|lcr}
\hline\hline       
\multicolumn{12}{c}{Ice, 1.35Myr}\\
Species & Abund. & Ratio & Species & Abund. & Ratio & Species & Abund. & Ratio & Species & Abund. & Ratio \\
\hline                    
C$_3$N & 2.4E-13 & 118 & C$_4$H$_4$ & 5.8E-12 & 2.59 & HC$_3$N & 2.1E-10 & 1.56 & H$_2$O & 1.3E-04 & 0.996 \\
C$_2$N & 1.6E-11 & 26.2 & CH$_3$C$_5$N & 4.9E-14 & 2.55 & S$_2$ & 3.4E-12 & 1.55 & N$_2$ & 2.6E-05 & 0.995 \\
HCCN & 1.7E-11 & 25.2 & NS & 4.7E-08 & 2.54 & CH$_3$CN & 2.5E-09 & 1.44 & CO & 8.9E-05 & 0.983 \\
C$_2$H$_2$N & 2.4E-11 & 12.3 & C$_3$S & 1.3E-11 & 2.34 & CH$_3$OCH$_3$ & 2.8E-12 & 1.37 & H$_2$S & 6.1E-07 & 0.978 \\
CN & 1.1E-09 & 5.52 & C$_6$H$_2$ & 4.5E-12 & 2.34 & C$_2$H$_5$OH & 2.8E-12 & 1.37 & HCOOCH$_3$ & 2.6E-13 & 0.971 \\
CS & 9.5E-09 & 4.97 & HC$_5$N & 1.1E-12 & 2.06 & C$_2$H$_6$ & 3.4E-08 & 1.23 & HCOOH & 8.1E-09 & 0.968 \\
HCS & 1.0E-08 & 4.94 & C$_5$H$_2$ & 4.6E-11 & 2.00 & H$_2$CCO & 1.6E-11 & 1.22 & H$_2$O$_2$ & 2.0E-06 & 0.927 \\
C$_2$H$_5$CN & 2.0E-12 & 4.67 & C$_4$N & 1.7E-13 & 2.00 & C$_2$H$_4$ & 1.3E-08 & 1.18 & O$_3$ & 9.4E-13 & 0.477 \\
H$_2$CS & 1.3E-07 & 4.29 & CH$_3$C$_3$N & 1.4E-11 & 1.87 & NH$_2$CHO & 4.8E-10 & 1.18 & SO$_2$ & 2.3E-10 & 0.427 \\
HNC & 3.8E-08 & 3.39 & C$_4$H$_2$ & 4.9E-10 & 1.77 & NH$_2$ & 1.0E-07 & 1.03 & O$_2$ & 2.3E-10 & 0.415 \\
C$_6$H$_6$ & 9.1E-14 & 3.10 & C$_2$H$_2$ & 4.2E-10 & 1.75 & CO$_2$ & 5.0E-05 & 1.02 & O$_2$H & 2.4E-10 & 0.405 \\
HC$_7$N & 1.0E-14 & 3.00 & CH$_3$CHO & 1.9E-13 & 1.67 & CH$_4$ & 1.4E-07 & 1.02 & SO & 6.6E-08 & 0.368 \\
CH$_3$C$_4$H & 8.2E-13 & 2.69 & HCN & 2.4E-08 & 1.66 & NH$_3$ & 1.9E-05 & 1.01 & C$_2$S & 2.4E-12 & 0.228 \\
\hline                  
\multicolumn{12}{c}{Gas, 1.388Myr}\\
H$_2$CCO & 1.8E-12 & 1.41 & N$_2$O & 1.1E-12 & 1.30 & H$_2$O & 1.3E-11 & 1.24 & HNC & 2.8E-10 & 1.12 \\
H$_2$CO & 2.7E-11 & 1.38 & CO$_2$ & 1.2E-08 & 1.29 & C$_2$H$_6$ & 1.8E-12 & 1.24 & HCN & 4.6E-10 & 1.12 \\
C$_3$ & 7.4E-12 & 1.30 & C$_3$H$_2$ & 1.2E-12 & 1.28 & CH$_4$ & 4.4E-08 & 1.18 & O$_2$ & 2.7E-10 & 0.42 \\
\hline                  
\end{tabular}
\end{table*}
\begin{figure}
  \vspace{-2cm}
  \hspace{-2cm}
  \includegraphics[width=14.0cm]{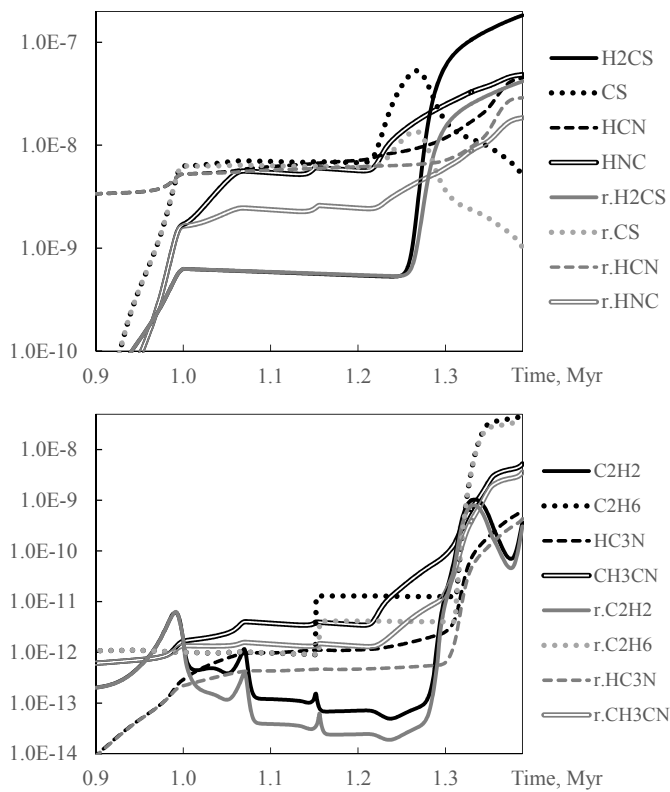}
  \vspace{-7.5cm}
  \caption{Abundance, relative to hydrogen, of selected ice molecules in model with CR-induced dissociation. The respective abundances from the reference model are shown for comparison.\label{att-crdis}}
\end{figure}
Deviations from the reference model because of direct CR-induced dissociation become observable already with the initiation of the subsurface ice layer at $t>0.7$Myr. Destruction of major species by CRs provides additional input of chemical radicals into the ice. This is unlike the mere redistribution the radicals generated by photodissociation, as for the whole-grain heating processes (Sects.~\ref{r-wdif} and \ref{r-warm}). This results in a pile-up of heavy radical species (Table~\ref{tab-crdis}). At 1.35Myr, this process has resulted in a 2\% decrease in CO abundance, when compared to the reference model. This has released significant amounts of C and O atoms that have combined in more complex species. This includes a 2\% increase in CO$_2$ abundance and an enhanced production of hydrogen-poor carbon chains, cyanopolyynes, and their associated radicals, HCN, HNC, HCCNH, H$_2$CS, and complex organic molecules. Figure~\ref{att-crdis} shows ice abundance for a selection of these species during the last stages of cloud evolution.

The abundances of hydrogen cyanide and isocyanide in ice are significantly increased because of the additional atomic and diatomic radicals produced by CR-induced dissociation. HNC is the more favored isomer because it readily forms from the C, NH, and NH$_2$ that arise from the dissociation of major ice constituents CO and NH$_3$. This does not affect the HNC:HCN gas-phase ratio at the temperature interval considered.

Regarding organic species, CR-induced dissociation produces (1) a notable increase in abundance for species containing carbon chains, and (2) a moderate increase in abundance for the oxygen-containing COMs. The first is associated with the enhanced production of C atoms from CO, while the second arises largely from the enhanced production of CH$_3$ radicals from methane. Because CO is much more abundant than CH$_4$, carbon chain chemistry is especially favored by dissociation processes. The O atoms, released by CO dissociation, are mostly absorbed into CO$_2$.

Thioaldehyde H$_2$CS is a relatively abundant ice molecule, whose abundance is significantly increased by CR-induced dissociation and is still growing at the end of the simulation run. This is because both of its formation pathways -- hydrogenation of CS and the reaction CH$_3$+S -- benefit from the increased radical abundances with the addition of CR-induced dissociation. The steep increase in H$_2$CS abundance, beginning at $t\approx1.25$Myr and $T\approx18$K, is because the 1000K barrier for the H+CS reaction can now be overcome. The increase is even more rapid for the model with CR-induced dissociation because of the additional production of C and S atoms from CO and H$_2$S, respectively. These atoms combine to form CS, which, in turn, is hydrogenated to H$_2$CS.

\subsection{The combined model}
\label{r-comb}
\begin{table*}
\caption{Calculation results for combined model -- molecular abundances, relative to hydrogen, for selected gas ice species and the abundance ratio (`R.'), relative to the reference model and the CR-induced dissociation (`crd') model.}
\label{tab-compl}
\centering
\begin{tabular}{lcrr|lcrr|lcrr}
\hline\hline       
\multicolumn{12}{c}{Ice, 1.35Myr}\\
Species & Abund. & R.ref. & R.crd. & Species & Abund. & R.ref. & R.crd. & Species & Abund. & R.ref. & R.crd. \\
\hline                    
C$_3$N & 2.2E-13 & 104 & 0.883 & C$_6$H$_2$ & 2.1E-12 & 1.10 & 0.470 & NO & 3.8E-15 & 0.997 & 1.413 \\
C$_2$N & 1.5E-11 & 23.8 & 0.911 & HC$_5$N & 5.7E-13 & 1.09 & 0.529 & H$_2$O & 1.3E-04 & 0.995 & 0.999 \\
HCCN & 1.6E-11 & 23.6 & 0.934 & CH$_3$NH & 1.4E-11 & 1.09 & 0.960 & CH$_2$NH$_2$ & 6.5E-10 & 0.992 & 0.921 \\
C$_2$H$_2$N & 2.3E-11 & 11.6 & 0.948 & CH$_3$C$_4$H & 3.3E-13 & 1.08 & 0.402 & HNO & 7.7E-08 & 0.991 & 1.006 \\
H$_2$C$_3$N & 4.4E-13 & 9.60 & 0.918 & C$_5$H$_2$ & 2.5E-11 & 1.08 & 0.540 & H$_2$C$_3$O & 3.4E-12 & 0.990 & 0.839 \\
C$_3$H$_3$N & 6.0E-13 & 7.19 & 0.944 & CH$_3$N & 2.7E-11 & 1.07 & 0.944 & CH$_4$O & 3.5E-10 & 0.989 & 0.993 \\
H$_4$C$_3$N & 6.7E-13 & 6.28 & 0.975 & HC$_2$NC & 5.3E-13 & 1.07 & 0.629 & CO & 8.9E-05 & 0.984 & 1.001 \\
OCN & 3.5E-12 & 5.64 & 0.908 & C$_4$H$_2$ & 2.9E-10 & 1.07 & 0.602 & NO$_2$ & 4.8E-11 & 0.982 & 1.008 \\
CN & 1.0E-09 & 5.24 & 0.950 & CH$_3$C$_3$N & 7.8E-12 & 1.06 & 0.567 & NH$_2$CN & 3.8E-11 & 0.978 & 0.993 \\
C$_2$H$_5$CN & 2.1E-12 & 4.86 & 1.041 & HC$_3$N & 1.4E-10 & 1.06 & 0.677 & H$_2$SIO & 1.9E-09 & 0.976 & 1.000 \\
HCS & 9.8E-09 & 4.72 & 0.956 & OCS & 1.6E-08 & 1.05 & 0.943 & H$_2$S & 6.1E-07 & 0.976 & 0.998 \\
CS & 9.0E-09 & 4.70 & 0.946 & C$_3$H$_2$ & 2.8E-09 & 1.05 & 0.668 & H$_2$S$_2$ & 5.8E-11 & 0.975 & 0.999 \\
H$_2$CS & 1.4E-07 & 4.34 & 1.010 & C$_3$H$_4$ & 7.0E-10 & 1.04 & 0.821 & H$_2$CO & 1.9E-08 & 0.974 & 0.997 \\
C$_4$H$_3$ & 1.5E-13 & 4.31 & 0.920 & C$_2$H$_4$ & 1.2E-08 & 1.04 & 0.877 & HCOOCH$_3$ & 2.6E-13 & 0.970 & 0.999 \\
HNC & 3.7E-08 & 3.31 & 0.977 & CH$_2$PH & 1.2E-13 & 1.03 & 0.496 & HS & 6.1E-07 & 0.969 & 1.002 \\
C$_2$H$_5$ & 2.1E-13 & 3.30 & 0.975 & CH$_3$NH$_2$ & 3.5E-08 & 1.03 & 0.921 & HCOOH & 8.1E-09 & 0.968 & 1.000 \\
C$_2$ & 6.1E-13 & 3.13 & 0.951 & CO$_2$ & 5.0E-05 & 1.03 & 1.001 & H$_2$ & 6.3E-15 & 0.967 & 1.255 \\
NS & 4.7E-08 & 2.56 & 1.005 & C$_2$H$_2$ & 2.5E-10 & 1.03 & 0.587 & NH$_2$ & 9.7E-08 & 0.964 & 0.941 \\
C$_3$S & 1.3E-11 & 2.33 & 0.995 & C$_2$H$_6$ & 2.8E-08 & 1.02 & 0.831 & HS$_2$ & 5.8E-11 & 0.964 & 0.999 \\
S$_2$ & 3.6E-12 & 1.64 & 1.058 & CH$_3$OCH$_3$ & 2.1E-12 & 1.02 & 0.745 & C$_3$H$_3$ & 1.4E-11 & 0.962 & 0.912 \\
HCN & 2.4E-08 & 1.63 & 0.980 & C$_2$H$_5$OH & 2.1E-12 & 1.02 & 0.745 & C$_3$H & 6.1E-13 & 0.953 & 0.857 \\
C$_4$H$_4$ & 3.3E-12 & 1.47 & 0.565 & CH$_4$ & 1.4E-07 & 1.02 & 1.002 & H$_2$O$_2$ & 2.0E-06 & 0.924 & 0.996 \\
NH$_2$CHO & 4.8E-10 & 1.18 & 1.000 & N$_2$H$_2$ & 1.4E-08 & 1.02 & 0.949 & O$_3$ & 8.6E-13 & 0.438 & 0.919 \\
CH$_3$CN & 2.0E-09 & 1.14 & 0.794 & HNCO & 9.5E-10 & 1.01 & 0.891 & SO$_2$ & 2.2E-10 & 0.405 & 0.949 \\
C$_8$H$_2$ & 1.8E-14 & 1.14 & 0.335 & NH & 5.3E-13 & 1.00 & 1.002 & O$_2$ & 2.2E-10 & 0.399 & 0.963 \\
CH$_3$CHO & 1.3E-13 & 1.13 & 0.680 & C$_2$H$_2$O & 1.3E-11 & 1.00 & 0.819 & O$_2$H & 2.3E-10 & 0.395 & 0.976 \\
HC$_7$N & 3.9E-15 & 1.13 & 0.377 & N$_2$ & 2.7E-05 & 0.999 & 1.004 & SO & 6.7E-08 & 0.371 & 1.007 \\
C$_7$H$_2$ & 1.8E-13 & 1.12 & 0.402 & NH$_3$ & 1.9E-05 & 0.998 & 0.990 & C$_2$S & 2.4E-12 & 0.229 & 1.006 \\
C$_6$H$_6$ & 3.3E-14 & 1.12 & 0.360 & N$_2$O & 2.7E-09 & 0.997 & 1.000 &  &  &  &  \\
\hline                  
\multicolumn{12}{c}{Gas, 1.388Myr}\\
C$_3$ & 5.9E-12 & 1.04 & 0.80 & H$_2$CO & 2.0E-11 & 1.02 & 0.74 & NH$_3$ & 6.8E-10 & 1.01 & 1.00 \\
C$_2$H$_6$ & 1.5E-12 & 1.03 & 0.83 & H$_2$O & 1.0E-11 & 1.02 & 0.82 & CO & 1.1E-05 & 0.99 & 0.99 \\
HNC & 2.5E-10 & 1.02 & 0.91 & CO$_2$ & 9.4E-09 & 1.01 & 0.78 & O$_2$ & 2.7E-10 & 0.43 & 1.02 \\
HCN & 4.2E-10 & 1.02 & 0.92 & CH$_4$ & 3.7E-08 & 1.01 & 0.86 &  &  &  &  \\
\hline                  
\end{tabular}
\end{table*}
\begin{figure}
  \vspace{-5cm}
  \hspace{-1cm}
  \includegraphics[width=11.0cm]{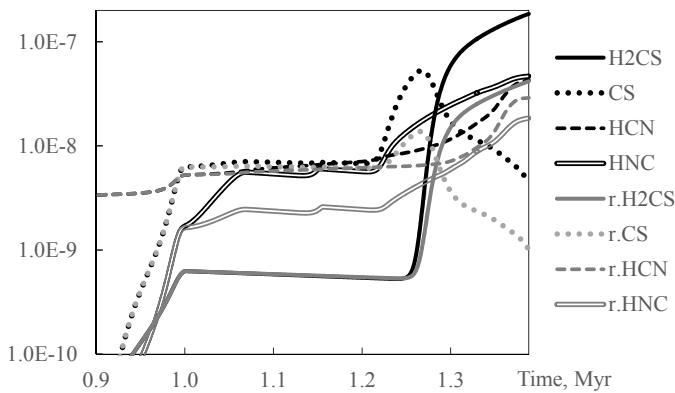}
  \vspace{-5cm}
  \caption{Abundance, relative to hydrogen, of selected ice molecules in the combined model. The respective abundances from the reference model are shown for comparison.\label{att-compl}}
\end{figure}
The complete model includes all three CR-induced processes considered in this paper: (1) CR-induced dissociation, (2) surface-mantle diffusion in ices affected by whole-grain heating, and (3) binary chemical reactions in such warmed-up ices. The complete model produces results mostly similar to those of the CR-induced dissociation model. Recall that, in the combined model, most of the CR-induced dissociation products are transferred to the warm phase (Sect.~\ref{crdissoc}). This leads to the overall result that more radicals combine into smaller molecules, most importantly, CH$_4$ and CO$_2$. The production of these species occurs via reactions with energy barriers involving a radical (C, CH$_2$, O, and OH) and species with high abundance in ice (H$_2$ and CO).

Nitrogen and sulfur radicals do not have such barriered reactions with major ice constituents. A small assortment of specific N and S compounds have their abundances higher than those calculated with the CR-induced dissociation model. This is because of energy barriers in synthesis reactions that involve at least one abundant radical. Although CR-induced dissociation favors the formation of COMs (Table~\ref{tab-crdis}), the abundances of these species are unchanged in the combined model, because any excess of the CH$_3$ radical, produced by CR-induced dissociation, is consumed in warmed-up ice by its barriered reaction with H$_2$. CH$_3$ is an essential ingredient in the synthesis of almost all complex organic species.

Table~\ref{tab-compl} shows the results of the combined model. A ratio of the combined model results against the reference model is provided. Dissociation is the main CR-induced process that influences ice composition. In the model, CR-induced dissociation has been reflected with a physically more precise methodology than warm-grain effects (see the introduction of Sect.~\ref{proc}). Because of this, the abundance ratio of the combined model versus CR-induced dissociation model is provided in Table~\ref{tab-compl}, too. Figure~\ref{att-compl} shows the abundance in the combined and reference models for a few stable ice species that are most significantly affected by the introduction of all three CR-induced effects considered in the present study. The complete model has gas phase abundances very similar to those of the reference model. This means that the enhanced rate of diffusion and reactions on the outer surface of warm grains is largely counterbalanced by CR-induced dissociation.

\section{Conclusions}
\label{concl}

Because of a complex grain temperature distribution, which is not explicitly considered here (Sect.~\ref{proc}), the investigation of diffusion and reactions on warm 70K grains is speculative and can only give insight into possible important processes. This is even more so because of the stochastic nature of Fe-CR hits to a grain, although the model presented here includes an attempt to tackle this issue. The promoted reactivity and molecule mobility on warm grains mostly result in enhanced production of carbon-chain compounds and other complex species. This agrees with the results of \citet{Reboussin14}, who investigated the effect of molecule diffusion on grains, induced by whole-grain heating. CR-induced mantle-surface diffusion via pores in the ice also shows similar results \citep{Kalvans14ba}.

The study of CR-induced dissociation, which is based on experimental data, shows that CR-induced dissociation significantly promotes the synthesis of carbon-chain compounds and COMs, and increases the HNC:HCN ratio. These results can probably be tested by observations of low-mass YSOs and detailed, object-specific chemical models.

In the combined model, the processes on grains warmed to 70K by CRs tend to reduce the effects caused by CR-induced dissociation. This is because several reactions with energy barriers that produce the major species CO$_2$ and CH$_4$ are made more effective. However, the radicals generated via dissociation can still serve as a source for COMs, which are observed in star-forming cores and not included in the current reaction network. Various carbon-chain molecules are among those with abundances that are most significantly increased by the CR-induced effects.

A direct comparison of the simulation results and observations with currently available data can be hard. The three modeled CR-induced processes do not affect the proportions of major species in interstellar ices -- H$_2$O, CO, CO$_2$, NH$_3$ \citep{Gibb04,Oberg11}. Additionally, these processes have little effect (less than 20\% abundance change, relative to the reference model) on observed interstellar gas-phase COMs \citep{Bacmann12,Cernicharo12,Marcelino05}. The formation of these species is believed to be associated with surface or subsurface ice chemistry \citep{Garrod08,Garrod13,Vasyunin13b,Kalvans14}.

Perhaps the most promising option for verifying the CR-induced effects is the gas-phase abundance of molecular oxygen. It is reduced by a factor of 2 in the complete model (protostellar phase, Table~\ref{tab-compl}), when molecule evaporation has already begun. In the interstellar medium O$_2$ has been detected in an environment that is subjected to outflows and shocks arising from nearby star-formation activity \citep{Larsson07,Liseau12,Chen14}. To the best of my knowledge, there have been no O$_2$ detections in quiescent medium or prestellar cores. O$_2$ has been found to have an abundance upper limit of $\approx10^{-8}$, relative to total H, in a low-mass protostar \citep{Yildiz13}. Thus a direct comparison of the presented calculation results and observations is currently not possible for the present study. The general increase in the abundances of organic species indicates that CR-induced processes in ice may play a role in the synthesis of these compounds.

It is interesting to note that the reduction of the O$_2$ gas phase abundance in the protostellar stage is mostly associated with CR-induced dissociation of species in the surface layer. Because the grains are gradually warmed up by the protostar, more oxygen atoms get included into CO$_2$, whose formation involves a small energy barrier. This is facilitated by the additional destruction of species (O$_2$) by CRs. It is possible that CR-induced dissociation contributes to not detecting O$_2$ in protostars.

\begin{acknowledgements}
I thank all three anonymous referees, who reviewed this manuscript at different stages. I am also grateful to A\&A editor Malcolm Walmsley for his contribution in improving the presentation. This research has made use of NASA's Astrophysics Data System.
\end{acknowledgements}

\bibliography{supra3}

\begin{thebibliography}{72}
\expandafter\ifx\csname natexlab\endcsname\relax\def\natexlab#1{#1}\fi

\bibitem[{{Accolla} {et~al.}(2011){Accolla}, {Congiu}, {Dulieu}, {Manic{\`o}},
  {Chaabouni}, {Matar}, {Mokrane}, {Lemaire}, \& {Pirronello}}]{Accolla11}
{Accolla}, M., {Congiu}, E., {Dulieu}, F., {et~al.} 2011, Physical Chemistry
  Chemical Physics (Incorporating Faraday Transactions), 13, 8037

\bibitem[{{Acharyya} {et~al.}(2011){Acharyya}, {Hassel}, \&
  {Herbst}}]{Acharyya11}
{Acharyya}, K., {Hassel}, G.~E., \& {Herbst}, E. 2011, \apj, 732, 73

\bibitem[{{Andersson} \& {van Dishoeck}(2008)}]{Andersson08}
{Andersson}, S. \& {van Dishoeck}, E.~F. 2008, \aap, 491, 907

\bibitem[{{Andrade} {et~al.}(2013){Andrade}, {de Barros}, {Pilling},
  {Domaracka}, {Rothard}, {Boduch}, \& {da Silveira}}]{Andrade13}
{Andrade}, D.~P.~P., {de Barros}, A.~L.~F., {Pilling}, S., {et~al.} 2013,
  \mnras, 430, 787

\bibitem[{{Bacmann} {et~al.}(2012){Bacmann}, {Taquet}, {Faure}, {Kahane}, \&
  {Ceccarelli}}]{Bacmann12}
{Bacmann}, A., {Taquet}, V., {Faure}, A., {Kahane}, C., \& {Ceccarelli}, C.
  2012, \aap, 541, L12

\bibitem[{{Bertin} {et~al.}(2013){Bertin}, {Fayolle}, {Romanzin}, {Poderoso},
  {Michaut}, {Philippe}, {Jeseck}, {{\"O}berg}, {Linnartz}, \&
  {Fillion}}]{Bertin13}
{Bertin}, M., {Fayolle}, E.~C., {Romanzin}, C., {et~al.} 2013, \apj, 779, 120

\bibitem[{{Brown} \& {Charnley}(1990)}]{Brown90}
{Brown}, P.~D. \& {Charnley}, S.~B. 1990, \mnras, 244, 432

\bibitem[{{Brown} {et~al.}(1988){Brown}, {Charnley}, \& {Millar}}]{Brown88}
{Brown}, P.~D., {Charnley}, S.~B., \& {Millar}, T.~J. 1988, \mnras, 231, 409

\bibitem[{{Brown} {et~al.}(1982){Brown}, {Augustyniak}, {Simmons},
  {Marcantonio}, {Lnazerotti}, {Johnson}, {Reimann}, {Foti}, \&
  {Pirronello}}]{Brown82}
{Brown}, W.~L., {Augustyniak}, W.~M., {Simmons}, E., {et~al.} 1982, Nuclear
  Instruments and Methods in Physics Research A, 198, 1

\bibitem[{{Caselli} {et~al.}(1998){Caselli}, {Hasegawa}, \&
  {Herbst}}]{Caselli98}
{Caselli}, P., {Hasegawa}, T.~I., \& {Herbst}, E. 1998, \apj, 495, 309

\bibitem[{{Cernicharo} {et~al.}(2012){Cernicharo}, {Marcelino}, {Roueff},
  {Gerin}, {Jim{\'e}nez-Escobar}, \& {Mu{\~n}oz Caro}}]{Cernicharo12}
{Cernicharo}, J., {Marcelino}, N., {Roueff}, E., {et~al.} 2012, \apjl, 759, L43

\bibitem[{{Chang} \& {Herbst}(2012)}]{Chang12}
{Chang}, Q. \& {Herbst}, E. 2012, \apj, 759, 147

\bibitem[{{Chen} {et~al.}(2014){Chen}, {Goldsmith}, {Viti}, {Snell}, {Lis},
  {Benz}, {Bergin}, {Black}, {Caselli}, {Encrenaz}, {Falgarone}, {Goicoechea},
  {Hjalmarson}, {Hollenbach}, {Kaufman}, {Melnick}, {Neufeld}, {Pagani}, {van
  der Tak}, {van Dishoeck}, \& {Yildiz}}]{Chen14}
{Chen}, J.-H., {Goldsmith}, P.~F., {Viti}, S., {et~al.} 2014, ArXiv e-prints

\bibitem[{{de Barros} {et~al.}(2011){de Barros}, {Domaracka}, {Andrade},
  {Boduch}, {Rothard}, \& {da Silveira}}]{deBarros11}
{de Barros}, A.~L.~F., {Domaracka}, A., {Andrade}, D.~P.~P., {et~al.} 2011,
  \mnras, 418, 1363

\bibitem[{{Du} \& {Parise}(2011)}]{Du11}
{Du}, F. \& {Parise}, B. 2011, \aap, 530, A131

\bibitem[{{Du} {et~al.}(2012){Du}, {Parise}, \& {Bergman}}]{Du12}
{Du}, F., {Parise}, B., \& {Bergman}, P. 2012, \aap, 538, A91

\bibitem[{{Fayolle} {et~al.}(2013){Fayolle}, {Bertin}, {Romanzin}, {Poderoso},
  {Philippe}, {Michaut}, {Jeseck}, {Linnartz}, {{\"O}berg}, \&
  {Fillion}}]{Fayolle13}
{Fayolle}, E.~C., {Bertin}, M., {Romanzin}, C., {et~al.} 2013, \aap, 556, A122

\bibitem[{{Fayolle} {et~al.}(2011){Fayolle}, {{\"O}berg}, {Cuppen}, {Visser},
  \& {Linnartz}}]{Fayolle11}
{Fayolle}, E.~C., {{\"O}berg}, K.~I., {Cuppen}, H.~M., {Visser}, R., \&
  {Linnartz}, H. 2011, \aap, 529, A74

\bibitem[{{Garrod}(2013)}]{Garrod13}
{Garrod}, R.~T. 2013, \apj, 765, 60

\bibitem[{{Garrod} \& {Herbst}(2006)}]{Garrod06}
{Garrod}, R.~T. \& {Herbst}, E. 2006, \aap, 457, 927

\bibitem[{{Garrod} \& {Pauly}(2011)}]{Garrod11}
{Garrod}, R.~T. \& {Pauly}, T. 2011, \apj, 735, 15

\bibitem[{{Garrod} {et~al.}(2007){Garrod}, {Wakelam}, \& {Herbst}}]{Garrod07}
{Garrod}, R.~T., {Wakelam}, V., \& {Herbst}, E. 2007, \aap, 467, 1103

\bibitem[{{Garrod} {et~al.}(2008){Garrod}, {Weaver}, \& {Herbst}}]{Garrod08}
{Garrod}, R.~T., {Weaver}, S.~L.~W., \& {Herbst}, E. 2008, \apj, 682, 283

\bibitem[{{Gerakines} {et~al.}(1996){Gerakines}, {Schutte}, \&
  {Ehrenfreund}}]{Gerakines96}
{Gerakines}, P.~A., {Schutte}, W.~A., \& {Ehrenfreund}, P. 1996, \aap, 312, 289

\bibitem[{{Gibb} {et~al.}(2004){Gibb}, {Whittet}, {Boogert}, \&
  {Tielens}}]{Gibb04}
{Gibb}, E.~L., {Whittet}, D.~C.~B., {Boogert}, A.~C.~A., \& {Tielens},
  A.~G.~G.~M. 2004, \apjs, 151, 35

\bibitem[{{Graninger} {et~al.}(2014){Graninger}, {Herbst}, {Oberg}, \&
  {Vasyunin}}]{Graninger14}
{Graninger}, D., {Herbst}, E., {Oberg}, K.~I., \& {Vasyunin}, A.~I. 2014, ArXiv
  e-prints

\bibitem[{{Hasegawa} \& {Herbst}(1993)}]{Hasegawa93a}
{Hasegawa}, T.~I. \& {Herbst}, E. 1993, \mnras, 261, 83

\bibitem[{{Hasegawa} {et~al.}(1992){Hasegawa}, {Herbst}, \&
  {Leung}}]{Hasegawa92}
{Hasegawa}, T.~I., {Herbst}, E., \& {Leung}, C.~M. 1992, \apjs, 82, 167

\bibitem[{{Iza} {et~al.}(2006){Iza}, {Farenzena}, {Jalowy}, {Groeneveld}, \&
  {da Silveira}}]{Iza06}
{Iza}, P., {Farenzena}, L.~S., {Jalowy}, T., {Groeneveld}, K.~O., \& {da
  Silveira}, E.~F. 2006, Nuclear Instruments and Methods in Physics Research B,
  245, 61

\bibitem[{{Kalv{\= a}ns}(2013)}]{Kalvans13b}
{Kalv{\= a}ns}, J. 2013, Space Research Review, 2, 15

\bibitem[{{Kalv{\= a}ns}(2014{\natexlab{a}})}]{Kalvans14ba}
{Kalv{\= a}ns}, J. 2014{\natexlab{a}}, Baltic Astronomy, 23, 137

\bibitem[{{Kalv{\= a}ns}(2014{\natexlab{b}})}]{Kalvans14}
{Kalv{\= a}ns}, J. 2014{\natexlab{b}}, submitted to ApJ

\bibitem[{{Kalv{\= a}ns} \& {Shmeld}(2010)}]{Kalvans10}
{Kalv{\= a}ns}, J. \& {Shmeld}, I. 2010, \aap, 521, A37

\bibitem[{{Kalv{\= a}ns} \& {Shmeld}(2013)}]{Kalvans13a}
{Kalv{\= a}ns}, J. \& {Shmeld}, I. 2013, \aap, 554, A111

\bibitem[{{Katz} {et~al.}(1999){Katz}, {Furman}, {Biham}, {Pirronello}, \&
  {Vidali}}]{Katz99}
{Katz}, N., {Furman}, I., {Biham}, O., {Pirronello}, V., \& {Vidali}, G. 1999,
  \apj, 522, 305

\bibitem[{{Larsson} {et~al.}(2007){Larsson}, {Liseau}, {Pagani}, {Bergman},
  {Bernath}, {Biver}, {Black}, {Booth}, {Buat}, {Crovisier}, {Curry},
  {Dahlgren}, {Encrenaz}, {Falgarone}, {Feldman}, {Fich}, {Flor{\'e}n},
  {Fredrixon}, {Frisk}, {Gahm}, {Gerin}, {Hagstr{\"o}m}, {Harju}, {Hasegawa},
  {Hjalmarson}, {Johansson}, {Justtanont}, {Klotz}, {Kyr{\"o}l{\"a}}, {Kwok},
  {Lecacheux}, {Liljestr{\"o}m}, {Llewellyn}, {Lundin}, {M{\'e}gie},
  {Mitchell}, {Murtagh}, {Nordh}, {Nyman}, {Olberg}, {Olofsson}, {Olofsson},
  {Olofsson}, {Persson}, {Plume}, {Rickman}, {Ristorcelli}, {Rydbeck},
  {Sandqvist}, {Sch{\'e}ele}, {Serra}, {Torchinsky}, {Tothill}, {Volk},
  {Wiklind}, {Wilson}, {Winnberg}, \& {Witt}}]{Larsson07}
{Larsson}, B., {Liseau}, R., {Pagani}, L., {et~al.} 2007, \aap, 466, 999

\bibitem[{{Lee} {et~al.}(1996){Lee}, {Herbst}, {Pineau des Forets}, {Roueff},
  \& {Le Bourlot}}]{Lee96}
{Lee}, H.-H., {Herbst}, E., {Pineau des Forets}, G., {Roueff}, E., \& {Le
  Bourlot}, J. 1996, \aap, 311, 690

\bibitem[{{Leger} {et~al.}(1985){Leger}, {Jura}, \& {Omont}}]{Leger85}
{Leger}, A., {Jura}, M., \& {Omont}, A. 1985, \aap, 144, 147

\bibitem[{{Li} {et~al.}(2013){Li}, {Heays}, {Visser}, {Ubachs}, {Lewis},
  {Gibson}, \& {van Dishoeck}}]{Li13}
{Li}, X., {Heays}, A.~N., {Visser}, R., {et~al.} 2013, \aap, 555, A14

\bibitem[{{Liseau} {et~al.}(2012){Liseau}, {Goldsmith}, {Larsson}, {Pagani},
  {Bergman}, {Le Bourlot}, {Bell}, {Benz}, {Bergin}, {Bjerkeli}, {Black},
  {Bruderer}, {Caselli}, {Caux}, {Chen}, {de Luca}, {Encrenaz}, {Falgarone},
  {Gerin}, {Goicoechea}, {Hjalmarson}, {Hollenbach}, {Justtanont}, {Kaufman},
  {Le Petit}, {Li}, {Lis}, {Melnick}, {Nagy}, {Olofsson}, {Olofsson}, {Roueff},
  {Sandqvist}, {Snell}, {van der Tak}, {van Dishoeck}, {Vastel}, {Viti}, \&
  {Y{\i}ld{\i}z}}]{Liseau12}
{Liseau}, R., {Goldsmith}, P.~F., {Larsson}, B., {et~al.} 2012, \aap, 541, A73

\bibitem[{{Marcelino} {et~al.}(2005){Marcelino}, {Cernicharo}, {Roueff},
  {Gerin}, \& {Mauersberger}}]{Marcelino05}
{Marcelino}, N., {Cernicharo}, J., {Roueff}, E., {Gerin}, M., \&
  {Mauersberger}, R. 2005, \apj, 620, 308

\bibitem[{{McElroy} {et~al.}(2013){McElroy}, {Walsh}, {Markwick}, {Cordiner},
  {Smith}, \& {Millar}}]{McElroy13}
{McElroy}, D., {Walsh}, C., {Markwick}, A.~J., {et~al.} 2013, \aap, 550, A36

\bibitem[{{Mej{\'{\i}}a} {et~al.}(2013){Mej{\'{\i}}a}, {de Barros}, {Bordalo},
  {da Silveira}, {Boduch}, {Domaracka}, \& {Rothard}}]{Mejia13}
{Mej{\'{\i}}a}, C.~F., {de Barros}, A.~L.~F., {Bordalo}, V., {et~al.} 2013,
  \mnras, 433, 2368

\bibitem[{{Meyer} {et~al.}(1998){Meyer}, {Drury}, \& {Ellison}}]{Meyer98}
{Meyer}, J.-P., {Drury}, L.~O., \& {Ellison}, D.~C. 1998, SSRv, 86, 179

\bibitem[{{Oba} {et~al.}(2009){Oba}, {Miyauchi}, {Hidaka}, {Chigai},
  {Watanabe}, \& {Kouchi}}]{Oba09}
{Oba}, Y., {Miyauchi}, N., {Hidaka}, H., {et~al.} 2009, \apj, 701, 464

\bibitem[{{{\"O}berg} {et~al.}(2011){{\"O}berg}, {Boogert}, {Pontoppidan}, {van
  den Broek}, {van Dishoeck}, {Bottinelli}, {Blake}, \& {Evans}}]{Oberg11}
{{\"O}berg}, K.~I., {Boogert}, A.~C.~A., {Pontoppidan}, K.~M., {et~al.} 2011,
  \apj, 740, 109

\bibitem[{{{\"O}berg} {et~al.}({2009a}){{\"O}berg}, {Linnartz}, {Visser}, \&
  {van Dishoeck}}]{Oberg09a}
{{\"O}berg}, K.~I., {Linnartz}, H., {Visser}, R., \& {van Dishoeck}, E.~F.
  {2009a}, \apj, 693, 1209

\bibitem[{{{\"O}berg} {et~al.}({2009b}){{\"O}berg}, {van Dishoeck}, \&
  {Linnartz}}]{Oberg09b}
{{\"O}berg}, K.~I., {van Dishoeck}, E.~F., \& {Linnartz}, H. {2009b}, \aap,
  496, 281

\bibitem[{{Palumbo}(2006)}]{Palumbo06}
{Palumbo}, M.~E. 2006, \aap, 453, 903

\bibitem[{{Pilling} {et~al.}(2010{\natexlab{a}}){Pilling}, {Seperuelo Duarte},
  {da Silveira}, {Balanzat}, {Rothard}, {Domaracka}, \& {Boduch}}]{Pilling10a}
{Pilling}, S., {Seperuelo Duarte}, E., {da Silveira}, E.~F., {et~al.}
  2010{\natexlab{a}}, \aap, 509, A87

\bibitem[{{Pilling} {et~al.}(2010{\natexlab{b}}){Pilling}, {Seperuelo Duarte},
  {Domaracka}, {Rothard}, {Boduch}, \& {da Silveira}}]{Pilling10b}
{Pilling}, S., {Seperuelo Duarte}, E., {Domaracka}, A., {et~al.}
  2010{\natexlab{b}}, \aap, 523, A77

\bibitem[{{Pirronello} {et~al.}(1982){Pirronello}, {Brown}, {Lanzerotti},
  {Marcantonio}, \& {Simmons}}]{Pirronello82}
{Pirronello}, V., {Brown}, W.~L., {Lanzerotti}, L.~J., {Marcantonio}, K.~J., \&
  {Simmons}, E.~H. 1982, \apj, 262, 636

\bibitem[{{Reboussin} {et~al.}(2014){Reboussin}, {Wakelam}, {Guilloteau}, \&
  {Hersant}}]{Reboussin14}
{Reboussin}, L., {Wakelam}, V., {Guilloteau}, S., \& {Hersant}, F. 2014,
  \mnras, 440, 3557

\bibitem[{{Roberts} {et~al.}(2007){Roberts}, {Rawlings}, {Viti}, \&
  {Williams}}]{Roberts07}
{Roberts}, J.~F., {Rawlings}, J.~M.~C., {Viti}, S., \& {Williams}, D.~A. 2007,
  \mnras, 382, 733

\bibitem[{{Roser} {et~al.}(2001){Roser}, {Vidali}, {Manic{\`o}}, \&
  {Pirronello}}]{Roser01}
{Roser}, J.~E., {Vidali}, G., {Manic{\`o}}, G., \& {Pirronello}, V. 2001,
  \apjl, 555, L61

\bibitem[{{Ruffle} \& {Herbst}(2001)}]{Ruffle01a}
{Ruffle}, D.~P. \& {Herbst}, E. 2001, \mnras, 322, 770

\bibitem[{{Savage} {et~al.}(1977){Savage}, {Bohlin}, {Drake}, \&
  {Budich}}]{Savage77}
{Savage}, B.~D., {Bohlin}, R.~C., {Drake}, J.~F., \& {Budich}, W. 1977, \apj,
  216, 291

\bibitem[{{Semenov} {et~al.}(2010){Semenov}, {Hersant}, {Wakelam}, {Dutrey},
  {Chapillon}, {Guilloteau}, {Henning}, {Launhardt}, {Pi{\'e}tu}, \&
  {Schreyer}}]{Semenov10}
{Semenov}, D., {Hersant}, F., {Wakelam}, V., {et~al.} 2010, \aap, 522, A42

\bibitem[{{Seperuelo Duarte} {et~al.}(2009){Seperuelo Duarte}, {Boduch},
  {Rothard}, {Been}, {Dartois}, {Farenzena}, \& {da Silveira}}]{Duarte09}
{Seperuelo Duarte}, E., {Boduch}, P., {Rothard}, H., {et~al.} 2009, \aap, 502,
  599

\bibitem[{{Seperuelo Duarte} {et~al.}(2010){Seperuelo Duarte}, {Domaracka},
  {Boduch}, {Rothard}, {Dartois}, \& {da Silveira}}]{Duarte10a}
{Seperuelo Duarte}, E., {Domaracka}, A., {Boduch}, P., {et~al.} 2010, \aap,
  512, A71

\bibitem[{{Shen} {et~al.}(2004){Shen}, {Greenberg}, {Schutte}, \& {van
  Dishoeck}}]{Shen04}
{Shen}, C.~J., {Greenberg}, J.~M., {Schutte}, W.~A., \& {van Dishoeck}, E.~F.
  2004, \aap, 415, 203

\bibitem[{{Taquet} {et~al.}(2012){Taquet}, {Ceccarelli}, \&
  {Kahane}}]{Taquet12}
{Taquet}, V., {Ceccarelli}, C., \& {Kahane}, C. 2012, \aap, 538, A42

\bibitem[{{Tielens}(2005)}]{Tielens05}
{Tielens}, A.~G.~G.~M. 2005, {The Physics and Chemistry of the Interstellar
  Medium} (Cambridge University Press)

\bibitem[{{van de Hulst}(1948)}]{vandeHulst48}
{van de Hulst}, H.~C. 1948, Harvard Observatory Monographs, 7, 73

\bibitem[{{Vasyunin} \& {Herbst}(2013a)}]{Vasyunin13a}
{Vasyunin}, A.~I. \& {Herbst}, E. 2013a, \apj, 762, 86

\bibitem[{{Vasyunin} \& {Herbst}(2013b)}]{Vasyunin13b}
{Vasyunin}, A.~I. \& {Herbst}, E. 2013b, \apj, 769, 34

\bibitem[{{Vasyunin} {et~al.}(2009){Vasyunin}, {Semenov}, {Wiebe}, \&
  {Henning}}]{Vasyunin09}
{Vasyunin}, A.~I., {Semenov}, D.~A., {Wiebe}, D.~S., \& {Henning}, T. 2009,
  \apj, 691, 1459

\bibitem[{{Webber} \& {Yushak}(1983)}]{Webber83}
{Webber}, W.~R. \& {Yushak}, S.~M. 1983, \apj, 275, 391

\bibitem[{{Whittet} {et~al.}(2007){Whittet}, {Shenoy}, {Bergin}, {Chiar},
  {Gerakines}, {Gibb}, {Melnick}, \& {Neufeld}}]{Whittet07}
{Whittet}, D.~C.~B., {Shenoy}, S.~S., {Bergin}, E.~A., {et~al.} 2007, \apj,
  655, 332

\bibitem[{{Willacy} {et~al.}(1994){Willacy}, {Williams}, \&
  {Duley}}]{Willacy94}
{Willacy}, K., {Williams}, D.~A., \& {Duley}, W.~W. 1994, \mnras, 267, 949

\bibitem[{{Y{\i}ld{\i}z} {et~al.}(2013){Y{\i}ld{\i}z}, {Acharyya}, {Goldsmith},
  {van Dishoeck}, {Melnick}, {Snell}, {Liseau}, {Chen}, {Pagani}, {Bergin},
  {Caselli}, {Herbst}, {Kristensen}, {Visser}, {Lis}, \& {Gerin}}]{Yildiz13}
{Y{\i}ld{\i}z}, U.~A., {Acharyya}, K., {Goldsmith}, P.~F., {et~al.} 2013, \aap,
  558, A58

\bibitem[{{Ziegler} {et~al.}(1985){Ziegler}, {Biersack}, \&
  {Littmark}}]{Ziegler85}
{Ziegler}, J.~F., {Biersack}, J.~P., \& {Littmark}, U. 1985, {The Stopping and
  Range of Ions in Solids} (Permagon Press, New York)

\end{thebibliography}
\bibliographystyle{aa}

\end{document}